\documentclass[journal,twoside,web]{ieeecolor}
\usepackage{generic}
\usepackage{cite}
\usepackage{amsmath,amssymb,amsfonts}
\usepackage{graphicx}
\usepackage{textcomp}

\usepackage{graphicx}  
\usepackage[english]{babel}
\usepackage{amsmath,amssymb}
\usepackage{autobreak}
\allowdisplaybreaks
\usepackage{mathrsfs}
\usepackage{color}
\usepackage{booktabs}
\usepackage{array}
\usepackage{svg}
\usepackage{siunitx} 
\usepackage{enumerate} 
\usepackage{graphicx}
\usepackage{caption}
\usepackage{multicol}  
\usepackage{multirow}  
\usepackage{flushend}
\usepackage{booktabs}
\usepackage{algorithm}
\usepackage[noend]{algpseudocode}
\usepackage{wrapfig}
\newtheorem{remark}{Remark} 
\newtheorem{lemma}{Lemma}

\newtheorem{theorem}{Theorem}

\newtheorem{assumption}{Assumption}
\newtheorem{problem}{Problem} 

\def\BibTeX{{\rm B\kern-.05em{\sc i\kern-.025em b}\kern-.08em
    T\kern-.1667em\lower.7ex\hbox{E}\kern-.125emX}}
\begin{document}
\title{Distributed Robust Model Predictive Control for
	Optimal Consensus of Unknown Multiagent Systems
	with Event-Triggered Communication}
\author{Shiyue Jia, Nan Bai, Dawei Shi, Ling Shi, \IEEEmembership{Fellow, IEEE}
\thanks{Shiyue Jia, Nan Bai and Ling Shi are with the Department of Electronic and Computer  Engineering (Ling Shi is also with the Department of Chemical and Biological Engineering and the Cheng Kar-Shun Robotics Institute),  Hong  Kong  University  of  Science  and Technology,  Clear  Water  Bay,  Kowloon,  Hong  Kong SAR (email: sjiaac@connect.ust.hk;
	eenanbai@ust.hk; eesling@ust.hk).}
\thanks{Dawei Shi is with MIIT Key Laboratory of Servo Motion System Drive and Control, School of Automation, Beijing Institute of Technology, Beijing, China (email: daweishi@bit.edu.cn).}}

\maketitle

\begin{abstract}
In this paper, we study the optimal consensus problem for multiagent systems with unknown dynamics. We propose a robust model predictive control (MPC) framework that jointly optimizes the control input and consensus sequences while enforcing state and input constraints. Within this framework, noisy offline data are used to construct nominal prediction models and quantify the associated uncertainty, which is explicitly incorporated into the robust MPC design. 
To solve the resulting optimization problem at each MPC update with reduced
inter-agent communication, we develop a distributed gradient-projection alternating direction method of multipliers (ADMM)
algorithm with an event-triggered information exchange mechanism and establish
its convergence to the optimal solution.
Furthermore, sufficient conditions are established to guarantee robust constraint satisfaction and practical consensus of the closed-loop multiagent system.
Numerical simulations demonstrate the effectiveness of the proposed framework.
\end{abstract}


\section{Introduction}
{C}{ooperative} control of multiagent systems (MASs) has attracted considerable attention in applications including mobile robotics
\cite{Bai2026LRA,Wang2025TAES},
networked sensing and state estimation \cite{Duan2024,Mo2024},
and intelligent energy systems
\cite{Yang2026,Wen2021}.
A fundamental coordination objective in such systems is consensus, where agents seek to reach agreement through local interactions
\cite{ferrari-trecate2009Model,johansson2008Decentralized}.
In many applications, however, reaching agreement alone is not sufficient, since the resulting consensus state and the transient behavior leading to it can have a substantial impact on overall system performance.
This has motivated the study of optimal consensus, where consensus is pursued together with an appropriate performance criterion
\cite{shi2013Reaching}.
Early studies typically optimized either the consensus state
\cite{qiu2016Distributed,xie2019Global}
or the transient control performance
\cite{yongcancao2010Optimal,movric2014Cooperative}.
More recently, these two objectives have been considered jointly
\cite{wang2019LQa,wang2022Accelerated},
allowing the agreement state and the transient control behavior to be coordinated within a simultaneous optimization framework.

The above optimal consensus approaches generally rely on accurate dynamical models of agents to predict the transient behavior associated with different consensus states and control inputs. In practice, however, real system dynamics may be unknown or only partially available and therefore need to be inferred from measured data, which are inevitably affected by noise and other uncertainties. Identification-based methods provide a natural way to address this issue by constructing nominal models from the collected data while quantifying the associated identification errors \cite{wu2023Distributed,parsi2020robust,lauricella2020Set,terzi2022Robust,lorenzen2019Robust}. These uncertainty bounds can then be explicitly incorporated into the subsequent control design rather than treating the identified models as exact. 
In particular, when state and input constraints must also be enforced, robust model predictive control (MPC) \cite{Mayne2005} and its distributed extensions \cite{wei2022Robust,wang2022Robust} provide a suitable framework for optimizing the consensus state and transient behavior based on the nominal dynamics while systematically accounting for uncertainties.

Within an MPC framework, however, the formulation of the optimal consensus problem differs from that of conventional tracking and coordination tasks. Specifically, in many existing MPC-based cooperative control schemes, the coordination reference is fixed before the online optimization procedure \cite{lyu2021Multivehicle,dunbar2006Distributed}.
Notice that such a predefined reference may be restrictive since different feasible consensus states may lead to substantially different transient trajectories and control efforts. It is therefore desirable to optimize the consensus state jointly with the control sequence, so that the eventual agreement point and the transient behavior can be coordinated under a unified performance objective. This joint design, together with local state and input constraints and the global consensus constraint, leads to a constrained coupled optimization problem that must be solved at each MPC update. For networked MASs, distributed optimization provides a natural means of solving this problem using only local computation and neighbor-to-neighbor information exchange \cite{bai2024Distributeda,bai2025Distributed,deng2025Distributed}.

It should be noted that solving the coupled constrained optimization problem in a distributed manner generally requires iterative exchanges of local primal and dual information among agents. Since this distributed optimization is performed repeatedly at each MPC update, frequent communication during the iterations may introduce considerable network overhead, particularly for bandwidth-limited or large-scale MASs \cite{liu2019CommunicationCensored,notarnicola2021Distributed}. 
Event-triggered communication offers a promising way to alleviate this burden
by transmitting local information only when prescribed triggering conditions
are satisfied, which has been incorporated into several distributed optimization
and networked coordination algorithms
\cite{liu2021ResourceAware,Wang2025CDC,Xu2024}.
For the problem considered here, however, the introduction of event-triggered communication is nontrivial due to the simultaneous optimization of the local control sequence and the coupled consensus state. Specifically, inter-agent communication is required for the consensus variable updates, while under event-triggered communication, the resulting communication errors may further affect the optimization of the local control sequence through the coupling between the two decision variables.
This calls for an event-triggered distributed optimization method that explicitly accounts for such couplings and establishes convergence under intermittent inter-agent communication.

Motivated by the above considerations, this paper studies constrained optimal consensus of MASs with unknown dynamics while reducing inter-agent communication. The problem requires addressing two key issues: \emph{how to ensure constraint satisfaction and closed-loop robustness when the MPC prediction model is identified from noisy data}, and \emph{how to reduce communication during distributed optimization while preserving convergence to the optimal solution}. 
Accordingly, the main contributions are summarized as follows.
\begin{enumerate}
	\item We propose a robust MPC framework  
	(Problem~\ref{prob:general_robust_mpc}) for optimal consensus of multiagent  
	systems with unknown dynamics, in which the control input and consensus  
	sequences are jointly optimized. Nominal prediction models and bounds on the  
	associated identification uncertainty are obtained from noisy data, and  
	the resulting uncertainty is explicitly incorporated into the MPC design to  
	achieve robust constraint satisfaction.
	
	\item We develop an event-triggered distributed gradient-projection alternating direction method of multipliers (ADMM) 
	algorithm for solving the resulting optimization problem at each 
	MPC update. An aggregated dual-variable formulation and an event-triggered 
	broadcasting mechanism are introduced to reduce inter-agent communication. 
	By explicitly accounting for the communication errors induced by intermittent 
	information exchange, convergence properties of the distributed algorithm are 
	established (Theorems~\ref{theorem:1} and~\ref{theorem:2}).
	
	\item We establish closed-loop guarantees for the proposed MPC framework under 
	identification-induced model uncertainty. Robust invariant sets and terminal 
	ingredients are constructed, and sufficient conditions are derived to guarantee 
	recursive feasibility, robust satisfaction of the state and input 
	constraints, and practical consensus for representative heterogeneous first-order 
	and second-order multiagent systems 
	(Theorems~\ref{theorem:first_order_closed_loop} 
	and~\ref{theorem:second_order_closed_loop}).
\end{enumerate}

\textbf{Notation:}
Let $\mathbb{R}^n$ and $\mathbb{R}^{m\times n}$ denote the spaces of
$n$-dimensional real column vectors and $m\times n$ real matrices,
respectively, while $I_n$ denotes the $n\times n$ identity matrix.
Let $\mathbb{N}=\{0,1,\ldots\}$ denote the set of nonnegative integers.
The notation
$\operatorname{col}\{x_1,\ldots,x_N\}
\triangleq[x_1^T,\ldots,x_N^T]^T$
denotes column-wise concatenation of vectors.
Let $\mathbf{1}$ and $\mathbf{0}$ denote the all-ones and
all-zeros column vectors of appropriate dimensions, respectively.
The symbol $x\otimes y$ denotes the Kronecker product.
The notation $\|\cdot\|$ denotes the Euclidean norm for vectors and
the induced $2$-norm for matrices.
For a matrix $P$, $P\succ0$ ($P\succeq0$) indicates that $P$ is
positive definite (positive semidefinite).
For symmetric $P$, $\underline{\lambda}(P)$ and
$\overline{\lambda}(P)$ denote its minimum and maximum eigenvalues,
respectively.
The weighted norm is defined as $\|x\|_P^2\triangleq x^TPx$,
and $\langle\cdot,\cdot\rangle$ denotes the inner product of two vectors.
The operator $\mathcal{P}_{\mathcal S}\{\cdot\}$ denotes the
Euclidean projection onto $\mathcal S$.
For sets $\mathcal X$ and $\mathcal Y$, their Minkowski sum and
Pontryagin difference are defined as
$\mathcal X\oplus\mathcal Y
\triangleq\{x+y\mid x\in\mathcal X,\,y\in\mathcal Y\}$
and
$\mathcal X\ominus\mathcal Y
\triangleq\{x\mid x+y\in\mathcal X,\,\forall y\in\mathcal Y\}$,
respectively.
The classes of $\mathcal K$-functions and
$\mathcal{KL}$-functions are denoted by $\mathcal K$ and
$\mathcal{KL}$, respectively.

\section{Preliminaries and Problem Formulation}
\subsection{System Model and Optimal Consensus Problem}
Consider a discrete-time multiagent system composed of $N$ heterogeneous agents with linear time-invariant dynamics
\begin{subequations}
	\label{eq:system_dynamics}
	\begin{align}
		x_i(k+1) &= A_i x_i(k)+B_i u_i(k), \label{eq:state_update}\\
		x_i(k) &\in \mathbb{X}_i, \label{eq:state_constraint}\\
		u_i(k) &\in \mathbb{U}_i,\quad i\in\{1,2,\dots,N\}, \label{eq:input_constraint}
	\end{align}
\end{subequations}
where $x_i(k)\in\mathbb{R}^n$ and $u_i(k)\in\mathbb{R}^m$ denote the state and control input of agent $i$, respectively, $A_i\in\mathbb{R}^{n\times n}$ and $B_i\in\mathbb{R}^{n\times m}$ are unknown constant system matrices, and $\mathbb{X}_i\subseteq\mathbb{R}^n$ and $\mathbb{U}_i\subseteq\mathbb{R}^m$ are compact convex sets.

The communication between the agents is modeled as an undirected graph $\mathcal{G} = (\mathscr{V}, \mathscr{E})$, where
$\mathscr{V}=
\{\boldsymbol{v}_1,\boldsymbol{v}_2,\dots,\boldsymbol{v}_N\}$ is the node set and
$\mathscr{E}\subseteq\mathscr{V}\times\mathscr{V}$ is the edge set.
An edge $(\boldsymbol{v}_i,\boldsymbol{v}_j)\in\mathscr{E}$ indicates that there is a communication channel between agent $i$ and agent $j$. The neighbor set of agent $i$ is defined as
$\mathcal{N}_i
\triangleq
\{j\in\{1,\dots,N\}\mid (\boldsymbol{v}_i,\boldsymbol{v}_j)\in\mathscr{E}\}.$
The adjacency matrix of $\mathcal{G}$ is denoted by
$\mathcal{A}=(a_{ij})\in\mathbb{R}^{N\times N}$, where
$a_{ii}=0$, $a_{ij}=a_{ji}=1$ if $(\boldsymbol{v}_i,\boldsymbol{v}_j)\in\mathscr{E}$,
and $a_{ij}=0$ otherwise. The Laplacian matrix
$\mathcal{L}=(l_{ij})\in\mathbb{R}^{N\times N}$ is defined by
$l_{ii}=\sum_{j=1}^{N}a_{ij}$ and $l_{ij}=-a_{ij}$ for $i\neq j$.

\begin{assumption}
	The communication graph $\mathcal{G}$ is fixed and connected.
	\label{ass:graph}
\end{assumption}

The control objective is to achieve optimal consensus among all agents under
state and input constraints.
Let $ X(k) \triangleq \operatorname{col}\{x_1(k),x_2(k),\ldots,x_N(k)\}$. Under Assumption~\ref{ass:graph}, the consensus condition $x_i(k)=x_j(k)$ for all $i,j\in\{1,\ldots,N\}$ can be equivalently characterized by $ (\mathcal L\otimes I_n)X(k)=\mathbf 0 $.
To facilitate controller design and characterize the agreement state, we introduce an auxiliary variable $z_i\in\mathbb Z_i$ for each agent, where $\mathbb Z_i\subseteq\mathbb X_i$ is a nonempty compact convex set representing the admissible reference states of agent $i$. For a given set of $z_i$, define $\boldsymbol z \triangleq \operatorname{col}\{z_1,z_2,\ldots,z_N\}$. 
Then, the admissible consensus set is given by
\begin{align}
	&\scalebox{0.88}{$\displaystyle
		\mathcal{Z}
		=
		\Big\{
		\boldsymbol z=\operatorname{col}\{z_1,z_2,\ldots,z_N\}
		\;\Big|\;
		(\mathcal{L}\otimes I_n)\boldsymbol z=\mathbf 0,\;
		z_i\in\mathbb{Z}_i
		\Big\}.
		$}
	\label{eq:z_constrain}
\end{align}
It should be emphasized that $z_i$ is not a prescribed consensus state that agent $i$ is required to track asymptotically. Instead, it will be optimized online to provide additional degrees of freedom for shaping the transient consensus performance. For a feasible $\boldsymbol z\in\mathcal Z$, define the corresponding consensus tracking deviation of agent $i$ as
\begin{equation}
	e_i(k)=x_i(k)-z_i,\qquad i\in\{1,2,\dots,N\}.
\end{equation}
In this work, we aim to jointly optimize both the consensus state
and the control input within an MPC framework, so that the transient consensus performance can be optimized while satisfying the state and input constraints. However, the true system matrices in \eqref{eq:system_dynamics} are unavailable for controller design, and only noisy input--state data collected from the agents are accessible. Consequently, the prediction models required by MPC must be constructed from data, while the resulting identification uncertainty needs to be explicitly accounted for to ensure reliable constraint satisfaction and closed-loop performance. To this end, we first construct nominal agent models and corresponding uncertainty bounds from the measured data, and then develop an identification-based robust MPC formulation.

\subsection{Model Identification and Robust Constraint Tightening}
\label{sec:Identification}

We first construct a nominal prediction model for each agent from offline input--state data corrupted by bounded state-measurement noise. Based on the collected data, nominal system matrices $\hat A_i$ and $\hat B_i$ are identified, yielding
\begin{equation}
	\hat x_i(k+1)
	=
	\hat A_i\hat x_i(k)+\hat B_i v_i(k),
	\label{eq:general_nominal_model}
\end{equation}
where $\hat x_i(k)\in\mathbb R^n$ and $v_i(k)\in\mathbb R^m$ denote the nominal system state and the corresponding control input, respectively.
Then, the discrepancy between the true and identified dynamics is represented as
\begin{equation}
	x_i(k+1)
	=
	\hat A_i x_i(k)+\hat B_i u_i(k)+\omega_i(k),
	\label{eq:general_uncertain_model}
\end{equation}
where $\omega_i(k)\in\mathbb R^n$ denotes the additive uncertainty induced
by the identification error. In particular, $\omega_i(k)
=
(A_i-\hat A_i)x_i(k)
+
(B_i-\hat B_i)u_i(k)$.
Under bounded measurement noise, sufficiently informative identification data,
and compact state and input constraint sets, the identification procedure
provides a finite uncertainty bound such that $
\|\omega_i(k)\|
\leq
\bar\omega_i,$
where $\bar\omega_i>0$ denotes the corresponding model uncertainty bound, as will be specified later.

To robustly accommodate the deviation between the actual and nominal
trajectories, define
$
\tilde e_i(k)
\triangleq
x_i(k)-\hat x_i(k),
$
and construct the  actual control input as
\begin{equation}
	u_i(k)
	=
	v_i(k)+K_i\tilde e_i(k),
	\label{eq:general_ancillary_control}
\end{equation}
where $K_i\in\mathbb R^{m\times n}$ is an ancillary feedback gain chosen
such that $\hat A_i+\hat B_iK_i$ is Schur stable.
Combining \eqref{eq:general_nominal_model},
\eqref{eq:general_uncertain_model}, and
\eqref{eq:general_ancillary_control} gives the error dynamics as
\begin{equation}
	\tilde e_i(k+1)
	=
	(\hat A_i+\hat B_iK_i)\tilde e_i(k)
	+
	\omega_i(k).
	\label{eq:error_dynamics}
\end{equation}
Let $
\mathbb W_i
\triangleq
\left\{
\omega_i\in\mathbb R^n
|
\|\omega_i\|\leq\bar\omega_i
\right\}
$
denote the admissible uncertainty set, and let
$\mathcal E_i\subseteq\mathbb R^n$ be a robust forward invariant set
for \eqref{eq:error_dynamics}, i.e.,
\begin{equation}
	(\hat A_i+\hat B_iK_i)\mathcal E_i
	\oplus
	\mathbb W_i
	\subseteq
	\mathcal E_i.
	\label{eq:rpi_condition}
\end{equation}
Hence, if $\tilde e_i(k)\in\mathcal E_i$, then
$\tilde e_i(k+\ell)\in\mathcal E_i$ for all $\ell\geq0$ under any
admissible uncertainty $\omega_i(k+\ell)\in\mathbb W_i$.

Accordingly, robust satisfaction of the original state and input constraints
can be enforced through the tightened nominal constraint sets $\hat{\mathbb X}_i
\triangleq
\mathbb X_i\ominus\mathcal E_i$, $\hat{\mathbb U}_i
\triangleq
\mathbb U_i\ominus K_i\mathcal E_i$.
Indeed, imposing
$
\hat x_i(k)\in\hat{\mathbb X}_i$,
$
v_i(k)\in\hat{\mathbb U}_i$
together with $\tilde e_i(k)\in\mathcal E_i$ guarantees
$x_i(k)\in\mathbb X_i$ and $u_i(k)\in\mathbb U_i$.
The detailed construction of the
robust invariant set $\mathcal E_i$ will be shown in
Section~\ref{sec:Robust Invariant Set Construction}.

\subsection{Identification-Based Robust MPC Formulation}
\label{sec:robust_mpc_formulation}

With the nominal prediction model and tightened constraint sets established, we next formulate the optimal consensus problem within an MPC framework. For a prediction horizon $\mathcal T$, define the nominal control sequence of agent $i$ as $\mathbf v_i(t_k) \triangleq \operatorname{col}\{ v_i(t_k|t_k),\ldots,v_i(t_k+\mathcal T-1|t_k) \}\in\mathbb{R}^{m\mathcal{T}}, $ and the corresponding predicted state sequence as $\hat{\mathbf{x}}_i(t_k) \triangleq \operatorname{col}\{ \hat x_i(t_k|t_k),\ldots,\hat x_i(t_k+\mathcal T|t_k) \}\in\mathbb{R}^{n(\mathcal T+1)}$.
At each MPC update time \(t_k\), let $\mathbf z_i(t_k) \triangleq \operatorname{col}\{ z_i(t_k|t_k),\ldots,z_i(t_k+\mathcal T|t_k) \}\in\mathbb{R}^{n(\mathcal{T}+1)}$ denote the auxiliary consensus sequence of agent $i$, and collect the consensus sequences of all agents as $Z(t_k) \triangleq \operatorname{col}\{ \mathbf z_1(t_k),\ldots, \mathbf z_N(t_k) \}.$
Similarly, define $V(t_k) \triangleq \operatorname{col}\{ \mathbf v_1(t_k),\ldots,\mathbf v_N(t_k) \}.$ 
Here, $V(t_k)$ and $Z(t_k)$ constitute the decision variables of the considered MPC problem.
To jointly optimize the transient consensus performance and control effort, consider the performance index 
\begin{equation} 
	J_{t_k} \big( V(t_k),Z(t_k) \big) = \sum_{i=1}^{N} J_{i,t_k} \big( \mathbf v_i(t_k),  \mathbf z_i(t_k) \big), 
	\label{eq:global_mpc_cost} 
\end{equation}
where 
\begin{align} 
	&J_{i,t_k} \big( \mathbf v_i(t_k), \mathbf z_i(t_k) \big)\nonumber  
	\\
	&\quad= \sum_{\ell=t_k}^{t_k+\mathcal T-1} 
	\left( 
	\left\| \hat x_i(\ell|t_k)-z_i(\ell|t_k) \right\|_{Q_i}^{2} 
	+ \left\| v_i(\ell|t_k) \right\|_{R_i}^{2} 
	\right) \nonumber\\ 
	&\qquad 
	+ \left\| \hat x_i(t_k+\mathcal T|t_k) 
	- z_i(t_k+\mathcal T|t_k) \right\|_{Q_{i \mathcal T }}^{2}, 
	\label{eq:local_mpc_cost} 
\end{align} 
with weighting matrices $Q_i\succ 0$, $R_i\succ 0$, and $Q_{i \mathcal T }\succ0$.

Let $\mathbb X_{iT}\subseteq\mathbb X_i$ and $\mathbb Z_{iT}\subseteq\mathbb Z_i$ denote nonempty compact convex terminal state and consensus sets, respectively. 
Based on the above ingredients, the robust MPC problem jointly optimizes the nominal control input $\mathbf v_i$ and the consensus sequence $\mathbf z_i$ while
accounting for the identification-induced model uncertainty through
constraint tightening. The resulting MPC optimization problem
is formulated as follows.

\begin{problem}
	\label{prob:general_robust_mpc}
	At each MPC update time $t_k$, for all $i\in\{1,\dots,N\}$, given the initial state $\hat x_i(t_k|t_k)=\hat x_i(t_k)$, solve the following optimization problem:
	\begin{subequations} 
		\begin{align} 
			\min_{V(t_k),\,Z(t_k)} \quad&
			J_{t_k}\big(V(t_k),Z(t_k)\big)\nonumber\\ 
			\mathrm{s.t.}\quad
			\hat x_i(\ell+1|t_k) 
			&= 
			\hat A_i\hat x_i(\ell|t_k) 
			+\hat B_i v_i(\ell|t_k),\label{x}\\ 
			v_i(\ell|t_k) 
			&\in 
			\mathbb U_i\ominus K_i\mathcal E_i,\label{U}\\ 
			\hat x_i(\ell|t_k) 
			&\in 
			\mathbb X_i\ominus\mathcal E_i,\label{X}\\ 
			z_i(\ell|t_k)
			&\in\mathbb Z_i,\quad 
			\ell\in[t_k,t_k+\mathcal T-1],\label{Z}\\ 
			\tilde{\mathcal{L}} Z(t_k) 
			&= \mathbf{0},\label{LZ}\\ 
			\hat x_i(t_k+\mathcal T|t_k) 
			&\in 
			\mathbb X_{iT}\ominus\mathcal E_i,\label{XT}\\ 
			z_i(t_k+\mathcal T|t_k)
			&\in\mathbb Z_{iT}.\label{ZT} 
		\end{align} 
	\end{subequations}
	where $\tilde{\mathcal{L}}=\mathcal{L}\otimes I_{n(\mathcal{T}+1)}$. 
	The constructions of terminal sets  $\mathbb X_{iT}$ and $\mathbb Z_{iT}$ used for the closed-loop analysis will be given in Section~\ref{sec:MPC}.
\end{problem}%
Problem~\ref{prob:general_robust_mpc} exhibits a partially coupled structure. Specifically, $\mathbf v_i(t_k)$ is a local decision variable of agent $i$, whereas $\mathbf z_i(t_k)$ is coupled with neighboring agents through the consensus constraint \eqref{LZ}. Meanwhile, $\mathbf v_i(t_k)$ and $\mathbf z_i(t_k)$ remain coupled through the local objective $J_{i,t_k}(\mathbf v_i(t_k),\mathbf z_i(t_k))$. 
In the following, we first develop a distributed optimization algorithm tailored to this structure in Section~\ref{sec:DPA}, with the aim of solving the MPC problem at each update time in a communication-efficient manner. We then analyze the closed-loop properties of the multiagent system under the proposed MPC framework in Section~\ref{sec:MPC}, including recursive feasibility, robust constraint satisfaction, and practical consensus under model uncertainty.

\section{Distributed Algorithm with Event-Triggered Communication}
\label{sec:DPA}

Within the proposed MPC framework, Problem~\ref{prob:general_robust_mpc} needs to be solved at every MPC update time $t_k$. The consensus constraint couples the consensus variables $\mathbf z_i$ of the agents. In the absence of a central coordinator, such coupling should be handled through iterative local updates and neighbor-to-neighbor information exchange. Since the optimization problem is repeatedly solved online, frequent communication during these iterations may lead to considerable communication overhead. To reduce unnecessary transmissions while preserving convergence to the optimal solution, we develop a distributed algorithm with event-triggered broadcasting.

For notational simplicity, the dependence on $t_k$ is omitted in this section whenever no confusion arises. Let $\bar{\mathbb{V}}_i$ denote the local feasible set of $\mathbf v_i$ determined by the nominal dynamics \eqref{x}, the tightened state and input constraints \eqref{U}--\eqref{X}, and the terminal state constraint \eqref{XT} in Problem~\ref{prob:general_robust_mpc}, and let $\bar{\mathbb{Z}}_i$ denote the feasible set of $\mathbf z_i$ determined by \eqref{Z} and \eqref{ZT}. Then, Problem~1 at each time $t_k$ can be compactly written as

\begin{problem}\label{prob:optimal_consensus}
	Optimization problem at each MPC update
	\begin{subequations}
		\label{eq:problem_constraints}
		\begin{align}
			\min_{V,Z} \quad & J(V,Z)=\sum_{i=1}^NJ_i(\mathbf v_i ,\mathbf z_i ) \nonumber\\
			\text{s.t. }
			&\mathbf v_i \in \bar{\mathbb{V}}_i, \quad
			\mathbf z_i \in \bar{\mathbb{Z}}_i, \label{eq:i_opt} \\
			&\tilde{\mathcal{L}} Z = \mathbf{0}. \label{eq:lap_opt} 
		\end{align}
	\end{subequations}
\end{problem}
\subsection{Algorithm Design}
From the compact formulation in Problem~\ref{prob:optimal_consensus},
the local constraints in \eqref{eq:i_opt} can be handled independently by individual agents, whereas \eqref{eq:lap_opt} constitutes the only coupling constraint among them. We therefore associate a dual variable with the consensus constraint and construct the augmented Lagrangian as follows
\begin{equation}
	L_\rho(V, Z, Y) = J(V, Z) + Y^TZ + \frac{\rho}{2} Z^T\tilde{\mathcal{L}}Z,
\end{equation}
where $Y = \tilde{\mathcal{L}}\Lambda$ is the aggregated dual variable, $\Lambda \triangleq \operatorname{col}\{\lambda_1, \lambda_2, \dots, \lambda_N\}$ denotes Lagrange multipliers corresponding to the consensus constraint \eqref{eq:lap_opt}, and $\rho > 0$ is the augmented Lagrangian parameter.

To mitigate the communication overhead, an event-triggered broadcasting mechanism is introduced into the network. Let $\mathcal{Q}=\{q\mid q\in\mathbb{N}\}$ denote the set of
optimization iteration indices, where the iteration counter $q$
serves as a global synchronizing clock for all agents. At each iteration $q$, we distinguish between the true local consensus variable $\mathbf{z}_{i}^q$ and the broadcast consensus variable $\hat{\mathbf{z}}_{i}^q$ for each agent $i$. Notably, all neighbors of agent $i$ maintain the same broadcast value
$\hat{\mathbf{z}}_i^q$. Let $\mathcal{Q}_i = \{q_i^l \mid l \in \mathbb{N}\} \subseteq \mathcal{Q}$ be the set of triggering iteration indices
for agent $i$. Assuming that all agents broadcast at initialization, i.e., $q_i^0=0$, each subsequent triggering instant $q_i^{l+1}$ obeys the following execution rule:
{\small\begin{equation}
		q_i^{l+1} = \min \left\{ q \in \mathcal{Q} \ \middle|\  q > q_i^l, \ \left\| \mathbf{z}_{i}^q - \hat{\mathbf{z}}_{i}^{q-1} \right\| \ge \sigma_i^{q} \right\},\nonumber
\end{equation}}%
where $q \ge 1$, and $\sigma_i^{q} \ge 0$ represents the dynamic triggering threshold. Accordingly, the broadcast variable $\hat{\mathbf{z}}_{i}^{q}$ inside the communication channel is defined as:
\begin{equation}
	\hat{\mathbf{z}}_i^{q+1}=
	\begin{cases}
		\mathbf{z}_i^{q+1}, & \text{if } q+1 \in \mathcal{Q}_i, \\[6pt]
		\hat{\mathbf{z}}_i^{q}, & \text{else}. 
	\end{cases}
	\label{eq:event_trigger_rule}
\end{equation}
Based on this mechanism, agent $i$ transmits its local variable $\mathbf{z}_i^{q+1}$ only when its deviation from the last broadcast value $\hat{\mathbf{z}}_i^q$ equals or exceeds the threshold. Consequently, the network communication error is bounded by the threshold sequence, ensuring that $\| \mathbf{z}_{i}^{q+1} - \hat{\mathbf{z}}_{i}^{q+1} \| \le \sigma_i^{q+1}$ holds for all $q+1 \in \mathcal{Q}$. For the triggering threshold, we make the following assumption.
\begin{assumption}
	The event-triggering thresholds $\{\sigma_i^q\}$ are non-negative and satisfy that $\sigma_i^q$ is nonincreasing for each $i$, and
	$
	\sum_{q=0}^{\infty} \sigma^q < \infty$, where $\sigma^q = \max_{i \in \{1,\dots,N\}} \sigma_i^q.
	$
	\label{ass:threshold}
\end{assumption}

Under this setup, the complete execution procedure of the proposed Event-Triggered Distributed Gradient Projection ADMM (ET-DGP-ADMM) is structured in Algorithm~\ref{alg:dhgpadmm}.
\begin{algorithm}[!htbp]
	\caption{ET-DGP-ADMM Algorithm.}
	\label{alg:dhgpadmm}
	\begin{algorithmic}[]
		\State \textbf{Initialize:} $q = 0$, $\mathbf{v}_i^0$,
		$\mathbf{z}_i^0$, $\hat{\mathbf{z}}_i^0 = \mathbf{z}_i^0$, $Y_i^0 = \mathbf{0}$, and termination tolerance
		$\varepsilon_i \geq 0$ for all $i$.
		\Repeat
		\begin{subequations}
			\begin{align}
				&\scalebox{0.9}{$\displaystyle
					\mathbf{v}_i^{q+1}
					=
					\arg\min_{\mathbf{v}_i \in \bar{\mathbb{V}}_i}
					\big\{
					J_{i}(\mathbf{v}_i, \mathbf{z}_i^q)
					+\tfrac{1}{2}(\mathbf{v}_i-\mathbf{v}_i^q)^T
					P_{vi}(\mathbf{v}_i-\mathbf{v}_i^q)
					\big\}
					$}
				\label{eq:update_u}\\
				&\scalebox{0.9}{$\displaystyle
					\mathbf{z}_i^{q+1}
					=
					\mathcal{P}_{\bar{\mathbb{Z}}_i}
					\bigg\{
					\mathbf{z}_i^q
					-p_{zi}\Big[
					\nabla_{\mathbf{z}_i}J_i(\mathbf{v}_i^{q+1},\mathbf{z}_i^q)
					+Y_i^q
					$}\nonumber\\
				&\scalebox{0.9}{$\displaystyle
					\qquad\qquad\qquad\quad
					+\rho\sum_{j\in\mathcal{N}_i}
					(\hat{\mathbf{z}}_i^q-\hat{\mathbf{z}}_j^q)
					\Big]\bigg\}
					$}
				\label{eq:update_z}
			\end{align}
		\end{subequations}
		\begin{align}
			&\scalebox{0.9}{$\displaystyle
				\hat{\mathbf{z}}_i^{q+1}=
				\begin{cases}
					\mathbf{z}_i^{q+1}, 
					& \text{if } 
					\left\|\mathbf{z}_i^{q+1}-\hat{\mathbf{z}}_i^q\right\|
					\ge \sigma_i^{q+1}, \\[6pt]
					\hat{\mathbf{z}}_i^q, 
					& \text{else}.
				\end{cases}
				$}
		\end{align}
		\State \textbf{communicate with neighbors} 
		$\hat{\mathbf{z}}_i^{q+1}$ \textbf{if} $\hat{\mathbf{z}}_i^{q+1}\neq \hat{\mathbf{z}}_i^{q}$
		\begin{align}
			&\scalebox{0.9}{$\displaystyle
				Y_i^{q+1}
				=
				Y_i^{q}
				+\rho\sum_{j\in\mathcal{N}_i}
				\big(
				\hat{\mathbf{z}}_i^{q+1}
				-
				\hat{\mathbf{z}}_j^{q+1}
				\big)
				$}
			\label{eq:update_y}
		\end{align}
		\State \textbf{set} $q = q + 1$.
		\Until{$| J_i(\mathbf{v}_i^q,\mathbf{z}_i^q)
			- J_i(\mathbf{v}_i^{q-1},\mathbf{z}_i^{q-1}) |
			\leq \varepsilon_i$}
		\State \textbf{output}: $\mathbf{v}_i^\star$, $\mathbf{z}_i^\star$.
	\end{algorithmic}
\end{algorithm}

\subsection{Convergence Analysis}
\label{sec:convergence_analysis}
In this subsection, we establish the theoretical guarantees for Algorithm~1. Due to the introduction of the event-triggered broadcasting mechanism, neighboring agents do not receive the exact current local consensus variables of each other, which introduces communication mismatches
into both the primal variable updates \eqref{eq:update_z} and the dual variable
updates \eqref{eq:update_y}. Nevertheless, the asymptotic convergence of the sequence generated by Algorithm~\ref{alg:dhgpadmm} is established in the following theorem.
\begin{theorem}
	Suppose that Assumptions \ref{ass:graph} and \ref{ass:threshold} hold,
	Problem~\ref{prob:optimal_consensus} is feasible, and
	$
	\operatorname{ri}(\bar{\mathbb Z})
	\cap
	\ker(\tilde{\mathcal L})
	\neq\emptyset,
	$
	where
	$
	\bar{\mathbb Z}
	\triangleq
	\left\{
	Z=\operatorname{col}\{\mathbf z_1,\ldots,\mathbf z_N\}
	\mid
	\mathbf z_i\in\bar{\mathbb Z}_i,\ \forall i\in\{1,\ldots,N\}
	\right\}.
	$
	Let $\{V^q, Z^q\}$ be the sequence generated by Algorithm~\ref{alg:dhgpadmm}. Then the sequence $\{V^q, Z^q\}$ converges to an optimal solution of Problem~\ref{prob:optimal_consensus}, if the auxiliary parameters satisfy the following conditions
	\begin{align*}
		P_{vi} \succ 0, \quad	0 < p_{zi} < \frac{1}{L_{hi} + 2\rho l_{ii}}, \quad\forall i \in \{1,2,\dots,N\},
	\end{align*}
	where $l_{ii}$ is the degree of agent $i$ in the communication graph, 
	and $L_{hi}$ is the Lipschitz constant of 
	$\nabla J_i(\mathbf v_i,\mathbf z_i)$, i.e., for any
	$[\mathbf v_i^p,\mathbf z_i^p]$ and $[\mathbf v_i^q,\mathbf z_i^q]$,  we have $\|\nabla J_i(\mathbf v_i^p,\mathbf z_i^p)
	-\nabla J_i(\mathbf v_i^q,\mathbf z_i^q)\|
	\le L_{hi}
	\left\|\operatorname{col}\{
	\mathbf v_i^p-\mathbf v_i^q,
	\mathbf z_i^p-\mathbf z_i^q
	\}\right\|.$
	\label{theorem:1}
\end{theorem}
\begin{proof}
	The proof is given in Appendix A.
\end{proof}

The convergence rate of Algorithm~\ref{alg:dhgpadmm} is further established in the following theorem.
\begin{theorem}
	Under the same assumptions and parameter conditions as in Theorem~\ref{theorem:1},
	the objective error satisfies 
	{\small\begin{align*}
			-\tfrac{\|\Lambda^\star\|}{2t(\gamma - \|\Lambda^\star\|)} \Gamma_t^2
			\le J(\bar{V}_t, \bar{Z}_t) - J(V^\star, Z^\star)
			\le \tfrac{1}{2t} \Gamma_t^2,
	\end{align*}}
	and the consensus error satisfies
	{\small\begin{align*}
			\|\tilde{\mathcal{L}} \bar{Z}_t\|
			\le \tfrac{1}{2t(\gamma - \|\Lambda^\star\|)} \Gamma_t^2,
	\end{align*}}%
	where $\bar{V}_t = \frac{1}{t} \sum_{q=1}^{t} V^q$ and $\bar{Z}_t = \frac{1}{t} \sum_{q=1}^{t} Z^q$ are the averaged iterates obtained by Algorithm~\ref{alg:dhgpadmm}; $\Gamma_t = \|V^\star - V^0\|_{P_V} + \|Z^\star - Z^0\|_{\Omega_Z} + c\gamma + \frac{2\sqrt{\lambda_b}}{\sqrt{3}} A_t$, with $\Omega_Z = P_Z^{-1} - \rho\tilde{\mathcal L}$, $\lambda_b = \min\{ \underline{\lambda}(P_V), \underline{\lambda}(\Omega_Z), \frac{1}{\overline{\lambda}(M)} \}$, $\lambda_a \triangleq \max\big\{1, 2\rho\overline{\lambda}(\tilde{\mathcal{L}})\big\}$, $A_t \triangleq \frac{3\lambda_a\sqrt{N}}{\lambda_b} \sum_{k=1}^{t} \sigma^{k-1}$, $c = \|\tilde{\mathcal L} \| \times \|M^{-1}\|^{1/2}$, and $\gamma > \|\Lambda^\star\|$.
	Since $A_t$ is uniformly bounded, both the objective error and the consensus error converge to zero at a rate of $O(1/t)$.
	\label{theorem:2}
\end{theorem}
\begin{proof}
	The proof is given in Appendix B.
\end{proof}

\section{Closed-loop Analysis of the Proposed MPC Framework}
\label{sec:MPC}
The preceding section proposes an efficient distributed algorithm to solve the optimization problem at each MPC update using the identified nominal model \eqref{eq:general_nominal_model}.
Recall that Section~\ref{sec:robust_mpc_formulation} formulated the identification-based robust MPC problem in terms of the identified nominal models $(\hat A_i,\hat B_i)$, the model uncertainty bounds $\bar\omega_i$, and the robust forward invariant sets $\mathcal E_i$. This section provides explicit constructions of these robustification ingredients from noisy offline data and subsequently investigates the closed-loop properties of the multiagent system within the proposed MPC framework. 

\subsection{Model Identification and Uncertainty Quantification}
\label{sec:datadriven_error}

This subsection provides the explicit identification and uncertainty
characterization underlying the robust MPC formulation in
Section~\ref{sec:robust_mpc_formulation}.

Recall the actual discrete-time heterogeneous agent dynamics introduced in
\eqref{eq:state_update}. The state measurements collected offline are modeled as
$
\tilde{x}_i(k)=x_i(k)+d_i(k)$,
where $d_i(k)\in\mathbb{R}^n$ is the bounded measurement noise satisfying
$\|d_i(k)\|\le\bar{d}_i$.
To identify the unknown matrix pair
$H_i^\star\triangleq[A_i,B_i]$, a sequence of inputs and noisy states is
collected offline over $L$ consecutive time steps and organized as $
X_{i-} \triangleq
[\tilde{x}_i(0),\tilde{x}_i(1),\dots,\tilde{x}_i(L-1)]$,
$U_{i-} \triangleq
[u_i(0),u_i(1),\dots,u_i(L-1)]$,
$X_{i+} \triangleq
[\tilde{x}_i(1),\tilde{x}_i(2),\dots,\tilde{x}_i(L)]$.
Define the stacked data matrix
$
\bar{\Phi}_i
\triangleq
\operatorname{col}\{X_{i-},U_{i-}\}
\in\mathbb{R}^{(n+m)\times L}.
$
The collected data are assumed to satisfy Assumption~\ref{ass:data_pe}.
\begin{assumption}
	\label{ass:data_pe}
	For each agent $i$, the data length satisfies $L\ge n+m$, and
	$
	\operatorname{rank}(\bar{\Phi}_i)=n+m
	$, which ensures that 
	$\bar{\Phi}_i\bar{\Phi}_i^T$ is nonsingular. 
\end{assumption}
%
Using the noisy state measurements, the collected data satisfy the regression
model $
X_{i+}=H_i^\star\bar{\Phi}_i+D_i$,
where $D_i$ denotes the lumped noise matrix induced by the measurement
noise. Accordingly, the nominal system matrices
$\hat H_i\triangleq[\hat A_i,\hat B_i]$ are obtained via least squares as
$
\hat H_i
=
X_{i+}\bar{\Phi}_i^T
(\bar{\Phi}_i\bar{\Phi}_i^T)^{-1}$.
The parameter estimation error matrix
$
\Delta_i
\triangleq
[\Delta_{A_i},\Delta_{B_i}]
=
\hat H_i-H_i^\star
=
D_i\bar{\Phi}_i^T
(\bar{\Phi}_i\bar{\Phi}_i^T)^{-1}
$
is bounded by
\begin{equation}
	\|\Delta_i\|
	\le
	\bar D_i
	\big(
	\underline{\lambda}
	(\bar{\Phi}_i\bar{\Phi}_i^T)
	\big)^{-\frac{1}{2}}
	\le
	\bar\delta_i,
	\label{eq:error_bound}
\end{equation}
where $\bar D_i>0$ satisfies $\|D_i\|\le\bar D_i$, and
$\bar\delta_i>0$ denotes an upper bound on the parameter estimation error.

Based on the least-squares estimate $\hat H_i=[\hat A_i,\hat B_i]$, the nominal
prediction model in \eqref{eq:general_nominal_model} is obtained. Moreover,
since $H_i^\star=\hat H_i-\Delta_i$, the actual dynamics can be rewritten in the
form of \eqref{eq:general_uncertain_model}, where
$
\omega_i(k)
=
-\Delta_i
\operatorname{col}\{x_i(k),u_i(k)\}
$
denotes the model uncertainty induced by the parameter estimation error.
Given \eqref{eq:error_bound} and the compactness of $\mathbb X_i$ and
$\mathbb U_i$, there exists a constant $\bar\omega_i>0$ satisfying
$
\|\omega_i(k)\|\le\bar\omega_i
$.
Hence, the uncertainty set $\mathbb W_i$ introduced in
Section~\ref{sec:Identification} can be constructed
from the offline data and the corresponding noise bound.

\subsection{Robust Invariant Set Construction}
\label{sec:Robust Invariant Set Construction}
We next provide an explicit construction of the robust forward invariant set $\mathcal E_i$ employed for constraint tightening in Problem~\ref{prob:general_robust_mpc}. 
Recall from \eqref{eq:error_dynamics} that the deviation between the actual and nominal trajectories evolves according to 
$\tilde e_i(k+1) = A_{K,i}\tilde e_i(k)+\omega_i(k)$, 
where $A_{K,i}\triangleq\hat A_i+\hat B_iK_i$ is Schur stable and 
$\omega_i(k)\in\mathbb W_i$. Consider the ellipsoidal set 
\begin{equation} 
	\mathcal E_i \triangleq \left\{ \tilde e_i\in\mathbb R^n \;\middle|\; \tilde e_i^T P_i\tilde e_i\leq\alpha_i \right\}, \label{eq:error_ellipsoidal_set} 
\end{equation} where $P_i\succ0$ and $\alpha_i>0$. The following result provides sufficient conditions under which $\mathcal E_i$ satisfies the robust invariance condition \eqref{eq:rpi_condition}.
\begin{lemma}[Robust Invariant Ellipsoid]
	\label{lem:invariant_ellipsoid}
	Consider the ellipsoidal set $\mathcal E_i$ defined in
	\eqref{eq:error_ellipsoidal_set}. Let $P_i\succ0$ satisfy
	\begin{equation}
		A_{K,i}^T P_i A_{K,i}-P_i+W_i\prec0,
		\label{eq:condition_p}
	\end{equation}
	for some $W_i\succ0$. If the scaling parameter $\alpha_i$ satisfies
	\begin{equation}
		\alpha_i \ge 
		\bar\lambda(P_i)d_{\omega,i}\bar\omega_i^2
		\left(c_{e,i}\right)^{-1},
		\label{eq:condition_alpha}
	\end{equation}
	where $0<\vartheta_i<1$,
	$c_{e,i}=(1-\vartheta_i)\underline{\lambda}(W_i)$,
	$d_{\omega,i}=\bar{\lambda}\big(P_i+\vartheta_i^{-1}\Pi_i\big)$, and
	$\Pi_i=P_iA_{K,i}W_i^{-1}A_{K,i}^TP_i$,
	then $\mathcal E_i$ is robustly forward invariant for the error dynamics
	\eqref{eq:error_dynamics}, i.e., if $\tilde e_i(k)\in\mathcal E_i$, then
	$\tilde e_i(k+1)\in\mathcal E_i$ for all
	$\|\omega_i(k)\|\le\bar\omega_i$.
\end{lemma}
\begin{proof}
	The proof is given in Appendix C.
\end{proof}

Lemma~\ref{lem:invariant_ellipsoid} provides an explicit construction of the set
$\mathcal E_i$ introduced in Section~\ref{sec:Identification}. Consequently,
$\tilde e_i(k)\in\mathcal E_i$ is preserved under all admissible
identification-induced uncertainties, and the tightened sets
$\mathbb X_i\ominus\mathcal E_i$ and
$\mathbb U_i\ominus K_i\mathcal E_i$
used in Problem~\ref{prob:general_robust_mpc} guarantee satisfaction of the original constraints.

\subsection{Robust MPC Implementation}
With the nominal models, uncertainty bounds, and robust invariant sets constructed above, we now describe the implementation of the proposed MPC framework. At each MPC update time $t_k$, Problem~\ref{prob:general_robust_mpc} is solved using Algorithm~\ref{alg:dhgpadmm} to obtain the optimal nominal control sequence $V^\star(t_k)$ and consensus variable $Z^\star(t_k)$. Let the control period satisfy $ 0<\Delta t\leq\mathcal T $. 
Only the first $\Delta t$ control moves of the optimized nominal control sequence are implemented before the optimization problem is solved again using updated state information.
For each agent $i$, the actual input applied during $\ell\in[t_k,t_{k+1})$ is\begin{equation} u_i(\ell ) = v_i^\star(\ell|t_k) + K_i \big( x_i(\ell)-\hat x_i^\star(\ell|t_k) \big), \label{apply}\end{equation} where $t_{k+1}=t_k+\Delta t$. The complete implementation is summarized in Algorithm~\ref{alg:data_informed_mpc}.

\begin{algorithm}
	\caption{Robust MPC Implementation.}
	\label{alg:data_informed_mpc}
	\begin{algorithmic}[1]
		\Require
		Prediction horizon $\mathcal T$, control period $\Delta t$,
		and algorithm parameters.
		
		\State Collect offline input-state data and identify the nominal
		models.
		
		\State Compute $\bar\omega_i$, select $K_i$, and construct
		$\mathcal E_i$.
		
		\State Initialize $t_0$ and the nominal states.
		
		\While{$t_k$ within the operation horizon}
		\State Set the initial state $\hat x_i(t_k|t_k)=\hat x_i(t_k)$ at $t_k$.
		
		\State Solve the finite-horizon optimization using
		Algorithm~\ref{alg:dhgpadmm} to obtain
		$V^\star(t_k)$ and $Z^\star(t_k)$.
		
		\State Apply the control input according to
		\eqref{apply}.
		
		\State Set $t_{k+1}=t_k+\Delta t$ and $k\leftarrow k+1$.
		
		\EndWhile
	\end{algorithmic}
\end{algorithm}

\subsection{Closed-Loop Analysis}
We now investigate the closed-loop properties of the proposed identification-based robust MPC framework. The general formulation in Problem~\ref{prob:general_robust_mpc} applies to heterogeneous linear systems, while explicit terminal ingredients and verifiable stability conditions depend on the particular agent dynamics. In this subsection, we specialize the general framework to representative first-order and second-order heterogeneous systems and establish the corresponding closed-loop guarantees.

For the dynamical models considered below, the general least-squares identification procedure in Section~\ref{sec:datadriven_error} is specialized by exploiting the known kinematic structure and estimating only the unknown scalar.
Moreover, define the aggregate state
deviation and model uncertainty as
$
\tilde e(k)
\triangleq
\operatorname{col}\{\tilde e_1(k),\ldots,\tilde e_N(k)\}
$
,
$
\omega(k)
\triangleq
\operatorname{col}\{\omega_1(k),\ldots,\omega_N(k)\}
$
,
$
P\triangleq\operatorname{blkdiag}\{P_1,\ldots,P_N\}
$,
$
c_e\triangleq\min_{i\in\{1,\ldots,N\}} c_{e,i}
$,
$
d_\omega\triangleq\max_{i\in\{1,\ldots,N\}} d_{\omega,i}
$,
and
$
\|\omega(k)\|\le\bar\omega
\triangleq
\sqrt{\sum_{i=1}^{N}\bar\omega_i^2}
$, which will be used for the subsequent closed-loop analysis.

\subsubsection{First-Order Systems}
Consider the first-order dynamics for each agent $i$:
\begin{subequations}
	\label{eq:first_order_system}
	\begin{align}
		& x_i(k+1) = x_i(k) + \kappa_i u_i(k), \label{eq:first_order_dynamics} \\
		& x_i(k) \in \mathbb{X}_i, \quad u_i(k) \in \mathbb{U}_i,  \quad i \in \{1,2,\dots,N\}, \label{eq:first_order_constraints}
	\end{align}
\end{subequations}
where $\kappa_i>0$ denotes the unknown input gain of agent $i$. 
\begin{remark}
	The system structure in \eqref{eq:first_order_system} can represent a broad class of practical systems with heterogeneous physical or actuation characteristics. For example, for a mechanical system governed by Newton's second law $F=ma$, if the velocity is taken as the state of interest, the corresponding discrete-time dynamics under a common sampling interval $\delta$ can be expressed in the form of \eqref{eq:first_order_dynamics}, with $\kappa_i=\frac{\delta}{m_i}$, where $m_i$ denotes the mass of agent $i$, which may not be precisely known during system modeling and controller design.
	\label{remark:firstorder}
\end{remark}

After the identification procedure, an estimate $\hat\kappa_i$ is obtained, leading to the nominal prediction model 
\begin{equation} \hat x_i(k+1) = \hat x_i(k)+\hat\kappa_i v_i(k). \label{eq:nominal_first_order} \end{equation} 
The corresponding identification error and induced model uncertainty are bounded according to the general construction developed in Section~\ref{sec:datadriven_error}. Hence, the robust invariant set $\mathcal E_i$ and tightened constraints introduced in Section~\ref{sec:Identification} apply directly. 

For the first-order system, the performance index \eqref{eq:local_mpc_cost} is specialized by choosing $Q_i=q_{s i}I$, $ R_i=r_{s i}I$, $ Q_{i\mathcal{T}}=p_{s i}I$, where $q_{s i}>0$, $r_{s i}>0$, and $p_{s i}>0$. Accordingly, Problem~\ref{prob:general_robust_mpc}, together with \eqref{eq:nominal_first_order}, defines the MPC problem for agents \eqref{eq:first_order_system}. The auxiliary consensus variable $z_i(\ell|t_k)$ remains optimized at every prediction stage and satisfies the consensus constraint \eqref{LZ}. To establish the closed-loop properties, we impose the following admissibility condition on the tightened state and input sets.
\begin{assumption}
	There exist constants $\gamma_i > 0$ and $\eta_i > 0$ such that $
	\hat{\mathcal{X}}_i := \{\hat x_i \mid \|\hat x_i\| \le \gamma_i \} \subseteq \mathbb{X}_i \ominus \mathcal{E}_i$,
	$\hat{\mathcal{V}}_i:=\{ v_i \mid \|v_i\| \le \eta_i \} \subseteq \mathbb{U}_i \ominus K_i\mathcal{E}_i$.
	\label{ass:constrant}
\end{assumption}

Then, the following theorem provides sufficient conditions for robust constraint satisfaction and closed-loop stability of the multiagent system \eqref{eq:first_order_system}. 

\begin{theorem} \label{theorem:first_order_closed_loop}
	Suppose that Problem~\ref{prob:general_robust_mpc}, specialized to
	\eqref{eq:first_order_system}, is feasible at the initial time $t_0$.
	Let $\mathcal E_i$ be the robust invariant set constructed in Lemma~\ref{lem:invariant_ellipsoid}, and choose the terminal sets as 
	$
	\mathbb Z_{iT} = \mathbb X_{iT}\ominus\mathcal E_i = \hat{\mathcal X}_i$. 
	Let the control period satisfy $\Delta t=1$. If, for every agent $i$, 
	\begin{align} 2\gamma_i-\hat\kappa_i\eta_i &\leq 0,\label{eq:first_order_feasibility_condition}
		\\ q_{s i}+\frac{r_{s i}}{\hat\kappa_i^2}-p_{s i} &\leq 0, \label{eq:first_order_stability_condition}
	\end{align} 
	and $\tilde e_i(t_0)\in\mathcal E_i$, then Problem~\ref{prob:general_robust_mpc} remains feasible at all subsequent MPC update times, and the actual states and inputs satisfy $x_i(k)\in\mathbb X_i$ and $u_i(k)\in\mathbb U_i$ for all admissible model uncertainties.
	Moreover, let $z^\star(t_k) \triangleq \operatorname{col}\{ z_1^\star(t_k|t_k),\ldots,z_N^\star(t_k|t_k) \}$ denote the first element of the optimized consensus sequences at time $t_k$.
	Then the closed-loop consensus tracking residual satisfies 
	\begin{equation}
		\|x(t_k)-z^\star(t_k)\|
		\leq
		\beta\bigl(
		\|(\hat x(t_0)-z^\star(t_0),\tilde e(t_0))\|,
		k
		\bigr)
		+r(\bar\omega).\nonumber
	\end{equation}
	where $\beta\in\mathcal{KL}$ and $r\in\mathcal K$. 
	%
	Furthermore, the consensus error of agents satisfies
	\begin{equation} 
		\|x_i(k)-x_j(k)\| 
		\leq 
		\|x_i(k)-z_i^\star(k|k)\| 
		+ 
		\|x_j(k)-z_j^\star(k|k)\|
		\nonumber 
	\end{equation} 
	due to the consensus constraint \eqref{LZ}, which implies that the actual multiagent system achieves practical consensus under bounded identification-induced uncertainty.
	
\end{theorem} 

\begin{proof}
	The proof is given in Appendix D. 
\end{proof}

\subsubsection{Second-Order Systems}
We next consider heterogeneous second-order systems that are described by 
\begin{subequations} \label{eq:second_order_system} 
	\begin{align} 
		&\begin{bmatrix} 
			p_i(k+1)\\ \dot p_i(k+1) \end{bmatrix} = \begin{bmatrix} I_n & \delta I_n\\ 0 & I_n \end{bmatrix} 
		\begin{bmatrix} p_i(k)\\ 
			\dot p_i(k) 
		\end{bmatrix} 
		+ \theta_i 
		\begin{bmatrix} 
			\frac{1}{2}\delta^2 I_n\\ 
			\delta I_n 
		\end{bmatrix} u_i(k), \label{eq:second_order_dynamics}\\ 
		&~p_i(k)\in\mathbb P_i,\quad \dot p_i(k)\in\mathbb V_i,\quad u_i(k)\in\mathbb U_i. \label{eq:second_order_constraints} \end{align} 
\end{subequations} 
\begin{remark}
	The second-order dynamics in \eqref{eq:second_order_system} arise in many practical applications. Here, $p_i$ and $\dot p_i$ denote the position and velocity of agent $i$, respectively, $u_i$ denotes the control input, and $\delta>0$ is the sampling interval. The scalar $\theta_i>0$ represents an unknown heterogeneous input gain. For example, for a mechanical system governed by Newton's second law, $\theta_i$ may correspond to the inverse mass $1/m_i$. Moreover, \eqref{eq:second_order_system} can be regarded as a special case of the general dynamics \eqref{eq:system_dynamics}, with
	$x_i=\operatorname{col}\{p_i,\dot p_i\}$ and
	$\mathbb X_i=\mathbb P_i\times\mathbb V_i$.
\end{remark}

Following Section~\ref{sec:datadriven_error}, an estimate
$\hat\theta_i$ is obtained from the offline data.
We assume that $\hat\theta_i\neq 0$. The resulting nominal model is
{\small\begin{equation*} \begin{bmatrix} \hat p_i(k+1)\\ \hat{\dot p}_i(k+1) \end{bmatrix} = \begin{bmatrix} I_n&\delta I_n\\ 0&I_n \end{bmatrix} \begin{bmatrix} \hat p_i(k)\\ \hat{\dot p}_i(k) \end{bmatrix} + \hat\theta_i \begin{bmatrix} \frac{1}{2}\delta^2I_n\\ \delta I_n \end{bmatrix} v_i(k). 
		\label{eq:nominal_second_order} 
\end{equation*}}%
For the second-order system, consensus is imposed on the position
variable $p_i$, while the velocity $\dot p_i$ is regulated toward zero.
Accordingly, the full-state auxiliary variable in the general
formulation of Problem~\ref{prob:general_robust_mpc} is specialized as $
\zeta_i(\ell|t_k)
\triangleq
\operatorname{col}\{
z_i(\ell|t_k),\mathbf 0
\}$,
where $z_i(\ell|t_k)\in\mathbb R^n$ denotes the consensus position
component. Thus, in this subsection, the full-state reference variable
in Problem~\ref{prob:general_robust_mpc} is represented by $\zeta_i$,
whereas $z_i$ is used exclusively for its position component.
The consensus constraint on $\zeta_i$ therefore reduces to consensus
of the position references $z_i$.
To recover the general performance index~\eqref{eq:local_mpc_cost}, choose
$Q_i = \operatorname{blkdiag} \{q_{x i}I_n,q_{v i}I_n\}$, $R_i = r_iI_n$, $Q_{i\mathcal{T}}=\operatorname{blkdiag} \{p_{x i}I_n,p_{v i}I_n\}$, where $q_{x i},q_{v i},r_i,p_{x i},p_{v i}>0$. The resulting stage cost therefore penalizes the position deviation $\hat p_i-z_i$, the nominal velocity $\hat{\dot p}_i$, and the nominal control effort. For the second-order system, write the actual-to-nominal deviation as $ \tilde e_i(k) = \operatorname{col} \left\{ e_i^p(k),e_i^v(k) \right\}$ with $e_i^p(k)=p_i(k)-\hat p_i(k)$ and $e_i^v(k)=\dot p_i(k)-\hat{\dot p}_i(k)$. 

Let the robust invariant ellipsoid constructed in Lemma~\ref{lem:invariant_ellipsoid} be written as 
\begin{equation} 
	\mathcal E_i = \left\{ \tilde e_i\in\mathbb R^{2n} \;\middle|\; 
	\tilde e_i^T P_i\tilde e_i\leq\alpha_i \right\}, 
	\label{eq:second_order_rpi} 
\end{equation} 
where
$P_i=\left[\begin{smallmatrix}
	P_{pp,i} & P_{pv,i}\\
	P_{pv,i}^T & P_{vv,i}
\end{smallmatrix}\right]\succ0$.
Its projections onto the position and velocity subspaces are
\begin{subequations} 
	\label{eq:projected_ellipsoids} 
	\begin{align} 
		\mathcal E_{p,i} &= \left\{ e_i^p\in\mathbb R^n \;\middle|\; 
		(e_i^p)^T 
		\left( P_{pp,i} - P_{pv,i} P_{vv,i}^{-1} P_{pv,i}^T \right) 
		e_i^p \leq\alpha_i \right\}, \\
		\mathcal E_{v,i} &= \left\{ e_i^v\in\mathbb R^n \;\middle|\; 
		(e_i^v)^T 
		\left( P_{vv,i} - P_{pv,i}^T P_{pp,i}^{-1} P_{pv,i} \right) 
		e_i^v \leq\alpha_i \right\}. 
	\end{align} 
\end{subequations}
Therefore, $\tilde e_i(k)\in\mathcal E_i$ implies $e_i^p(k)\in\mathcal E_{p,i}$ and $e_i^v(k)\in\mathcal E_{v,i}$. 
Since
\(
\mathbb X_i=\mathbb P_i\times\mathbb V_i,
\)
the tightened full-state constraint in
Problem~\ref{prob:general_robust_mpc} can be represented componentwise as
\(
\hat p_i\in\mathbb P_i\ominus\mathcal E_{p,i}\) and
\(\hat{\dot p}_i\in\mathbb V_i\ominus\mathcal E_{v,i}.
\)
We impose the following assumption for the terminal construction. 

\begin{assumption} \label{ass:second_order_constraint} For each agent $i$, there exist constants $\bar\gamma_i^p>0$, $\bar\gamma_i^v>0$, and $\bar\eta_i>0$ such that 
	$
	\hat{\mathcal X}_i^p \triangleq\{ \hat p_i\in\mathbb R^n | \|\hat p_i\|\leq\bar\gamma_i^p \} \subseteq \mathbb P_i\ominus\mathcal E_{p,i}$,
	$\hat{\mathcal P}_i^v \triangleq \{ \hat{\dot p}_i\in\mathbb R^n | \|\hat{\dot p}_i\|\leq\bar\gamma_i^v \} \subseteq \mathbb V_i\ominus\mathcal E_{v,i}$, 
	$\hat{\mathcal V}_i \triangleq \{ v_i\in\mathbb R^n |\|v_i\|\leq\bar\eta_i \} \subseteq \mathbb U_i\ominus K_i\mathcal E_i$. 
\end{assumption} 

The closed-loop properties of the second-order multiagent systems are summarized in the following theorem.

\begin{theorem}
	\label{theorem:second_order_closed_loop}
	Suppose that Problem~\ref{prob:general_robust_mpc}, specialized to
	\eqref{eq:second_order_system}, is feasible at the initial time $t_0$.
	Let $\mathcal E_i$ be the robust invariant ellipsoid in
	\eqref{eq:second_order_rpi}, with projections $\mathcal E_{p,i}$ and
	$\mathcal E_{v,i}$ given by \eqref{eq:projected_ellipsoids}.
	Choose the terminal sets as
	$
	\mathbb Z_{iT}
	=
	\mathbb P_{iT}\ominus\mathcal E_{p,i}
	=
	\{
	\hat p_i
	\mid
	\|\hat p_i\|\leq\gamma_i^p
	\}$,
	$
	\mathbb V_{iT}\ominus\mathcal E_{v,i}
	=
	\{
	\hat{\dot p}_i
	\mid
	\|\hat{\dot p}_i\|\leq\gamma_i^v
	\}$.
	Suppose that, for every agent $i$, 
	\begin{align} \gamma_i^p+\frac{1}{4}\delta\gamma_i^v &\leq \bar\gamma_i^p, \label{eq:second_order_state_condition}\\ 4\gamma_i^p+3\delta\gamma_i^v &\leq 2\delta^2|\hat\theta_i|\bar\eta_i, \label{eq:second_order_input_condition}\\ 2\gamma_i^p+\frac{1}{2}\delta\gamma_i^v &\leq \delta\bar\gamma_i^v, \label{eq:second_order_velocity_condition} 
	\end{align} 
	and the weighting parameters satisfy 
	\begin{align} \frac{4r_i}{\hat\theta_i^2\delta^4} +\frac{3q_{x i}}{2} +\frac{2q_{v i}}{\delta^2} -p_{x i} &\leq0, \label{eq:second_order_weight_condition_1}\\ \frac{5r_i}{\hat\theta_i^2\delta^2} +\frac{3q_{v i}}{2} +\frac{q_{x i}\delta^2}{8} -p_{v i} &\leq0. \label{eq:second_order_weight_condition_2} 
	\end{align}
	If $2\leq\Delta t\leq\mathcal T$ and $\tilde e_i(t_0)\in\mathcal E_i$ for all $i$, then Problem~\ref{prob:general_robust_mpc} remains feasible at all subsequent MPC update times, and the actual closed-loop system satisfies the original state and input constraints for all admissible model uncertainties. 
	Moreover, for each $\tau\in[t_k,t_{k+1})$, define the implemented
	consensus reference as
	$
	z^\star(\tau|t_k)
	\triangleq
	\operatorname{col}\{
	z_1^\star(\tau|t_k),\ldots,
	z_N^\star(\tau|t_k)
	\}.
	$
	Then, at the MPC update instants, the closed-loop position tracking residual satisfies $
	\|p(t_k)-z^\star(t_k|t_k)\|
	\leq
	\bar\beta\bigl(
	\|(\hat p(t_0)-z^\star(t_0|t_0),
	\hat{\dot p}(t_0),
	\tilde e(t_0))\|,
	t_k-t_0
	\bigr)+\bar r(\bar\omega)$,
	where $\bar\beta\in\mathcal{KL}$ and $\bar r\in\mathcal K$.
	Furthermore, 
	we have
	\begin{equation}
		\|p_i(\tau)-p_j(\tau)\|
		\leq
		\|p_i(\tau)-z_i^\star(\tau|t_k)\|
		+
		\|p_j(\tau)-z_j^\star(\tau|t_k)\|
		\nonumber
	\end{equation}
	for all $\tau\in[t_k,t_{k+1})$ and $k\geq0$.
	Therefore, the position disagreement remains ultimately bounded
	for all time instants under bounded identification-induced uncertainty,
	and the actual positions achieve practical consensus.
	
\end{theorem}

\begin{proof}
The proof is given in Appendix E. 	
\end{proof}

\section{Simulation Examples}
This section presents numerical experiments to verify the theoretical results and evaluate the practical performance of the proposed method under different scenarios.
\subsection{ET-DGP-ADMM Within a Finite Horizon}

\begin{figure}[!htbp]
	\centering
	\includegraphics[width=0.35\linewidth]{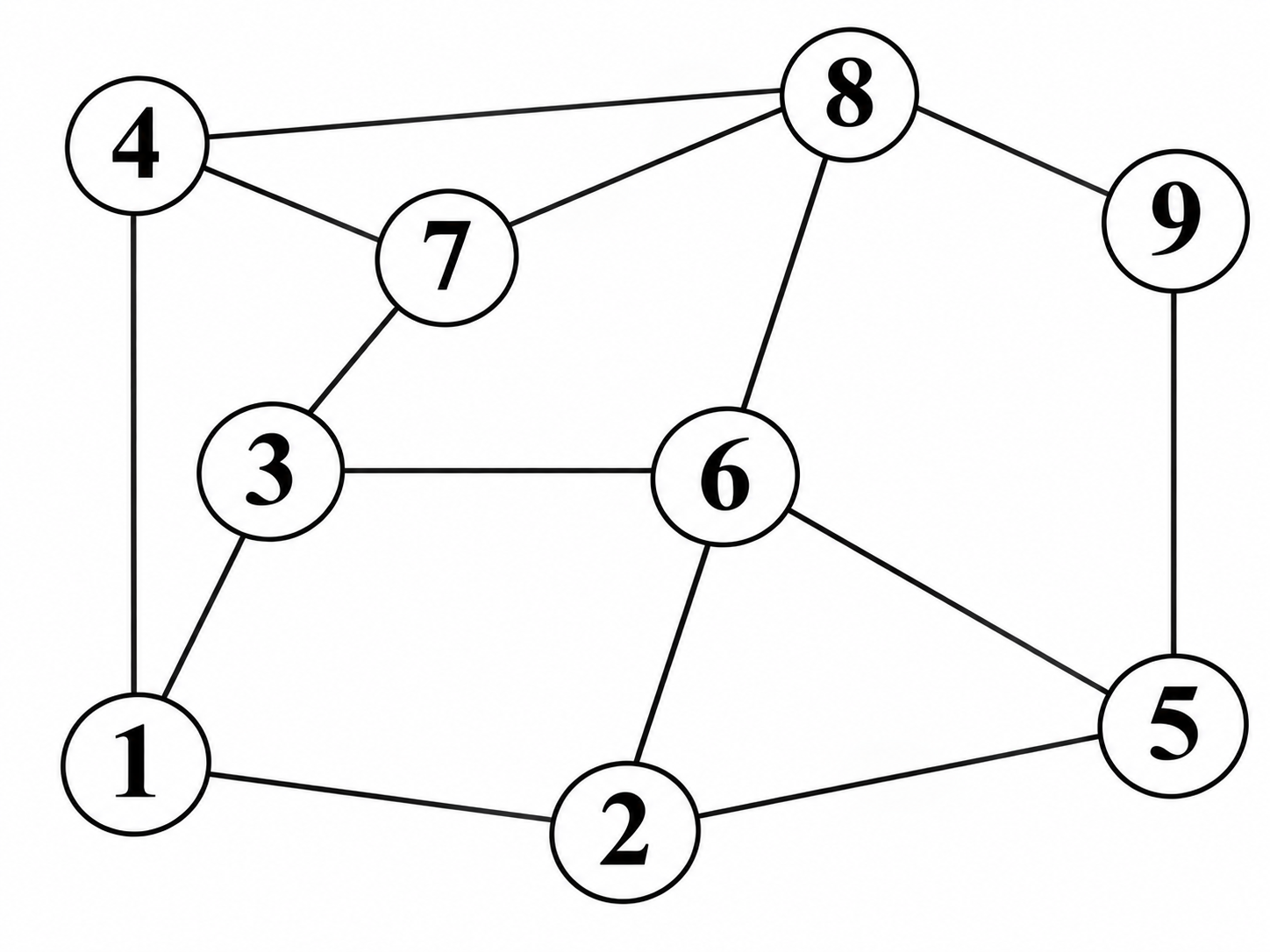}
	\caption{Communication topology of the nine agents.}
	\label{fig:communication_topology}
\end{figure}

Consider a system consisting of nine agents interconnected through the
communication topology shown in Figure~\ref{fig:communication_topology}, with the heterogeneous dynamics
matrices
{\small\begin{align*}
		A_i&=
		\begin{bmatrix}
			1+0.1\Delta_i & 1+0.1\Delta_i & 0\\
			0 & 1+0.1\Delta_i & 1+0.1\Delta_i\\
			0 & -1+0.1\Delta_i & -1+0.1\Delta_i
		\end{bmatrix},
		\\
		B_i&=
		\begin{bmatrix}
			1+0.1\Delta_i & 0\\
			0 & 1+0.1\Delta_i\\
			1+0.1\Delta_i & 1+0.1\Delta_i
		\end{bmatrix}, \quad i\in\{1,2,\ldots,9\}.
\end{align*}}%
Here, \(\Delta_i\) is randomly generated from the interval \([-1,1]\).
The resulting models are heterogeneous, with the perturbed parameters varying within $\pm10\%$ of their nominal values.

The proposed algorithm is implemented with the finite horizon $\mathcal T=15$. For all \(i\in\{1,2,\ldots,9\}\), all three state
components participate in the consensus objective, and the weighting
matrices are selected as \(Q_i=I_3\), \(Q_{i\mathcal T}=I_3\), and
\(R_i=0.1I_2\).
The state and input constraint sets are given by
\(\mathbb{X}_i=\{x_i\in\mathbb{R}^3\mid
\underline{x}\leq x_i\leq\overline{x}\}\) and
\(\mathbb{U}_i=\{u_i\in\mathbb{R}^2\mid
\underline{u}\leq u_i\leq\overline{u}\}\), respectively, where
\(\underline{x}=[-4,-2,-4]^{T}\),
\(\overline{x}=[4,2,4]^{T}\),
\(\underline{u}=[-3,-3]^{T}\), and
\(\overline{u}=[3,3]^{T}\). The consensus constraint set is
selected as \(\mathbb{Z}_i=\mathbb{X}_i\).
The algorithmic parameters are selected as
\(P_{vi}=0.1I_{30}\) and \(p_{zi}=0.01\).
The event-triggering threshold is defined by
\(\sigma_i^q=\sigma_i^0\beta^q\), with
\(\sigma_i^0=8\) and \(\beta=0.95\), so that the resulting threshold
sequence satisfies the summability condition. The stopping
tolerance is set to \(\varepsilon_i=10^{-5}\).
\begin{figure}[!htbp]
	\centering
	\includegraphics[width=0.8\linewidth]{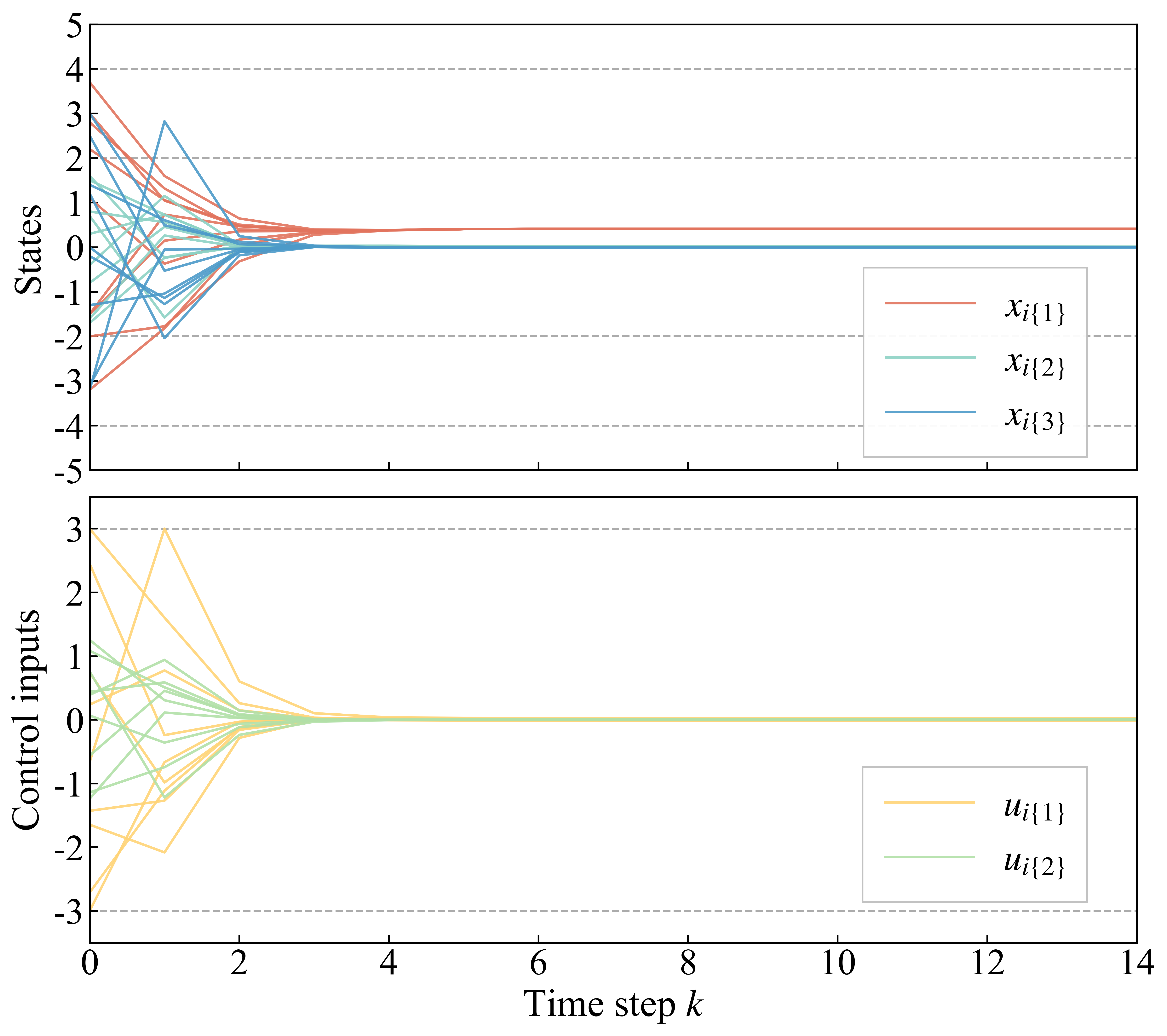}
	\caption{State and control-input trajectories of agents.}
	\label{fig:state_control}
\end{figure}
\begin{figure}[!htbp]
	\centering
	\includegraphics[width=0.85\linewidth]{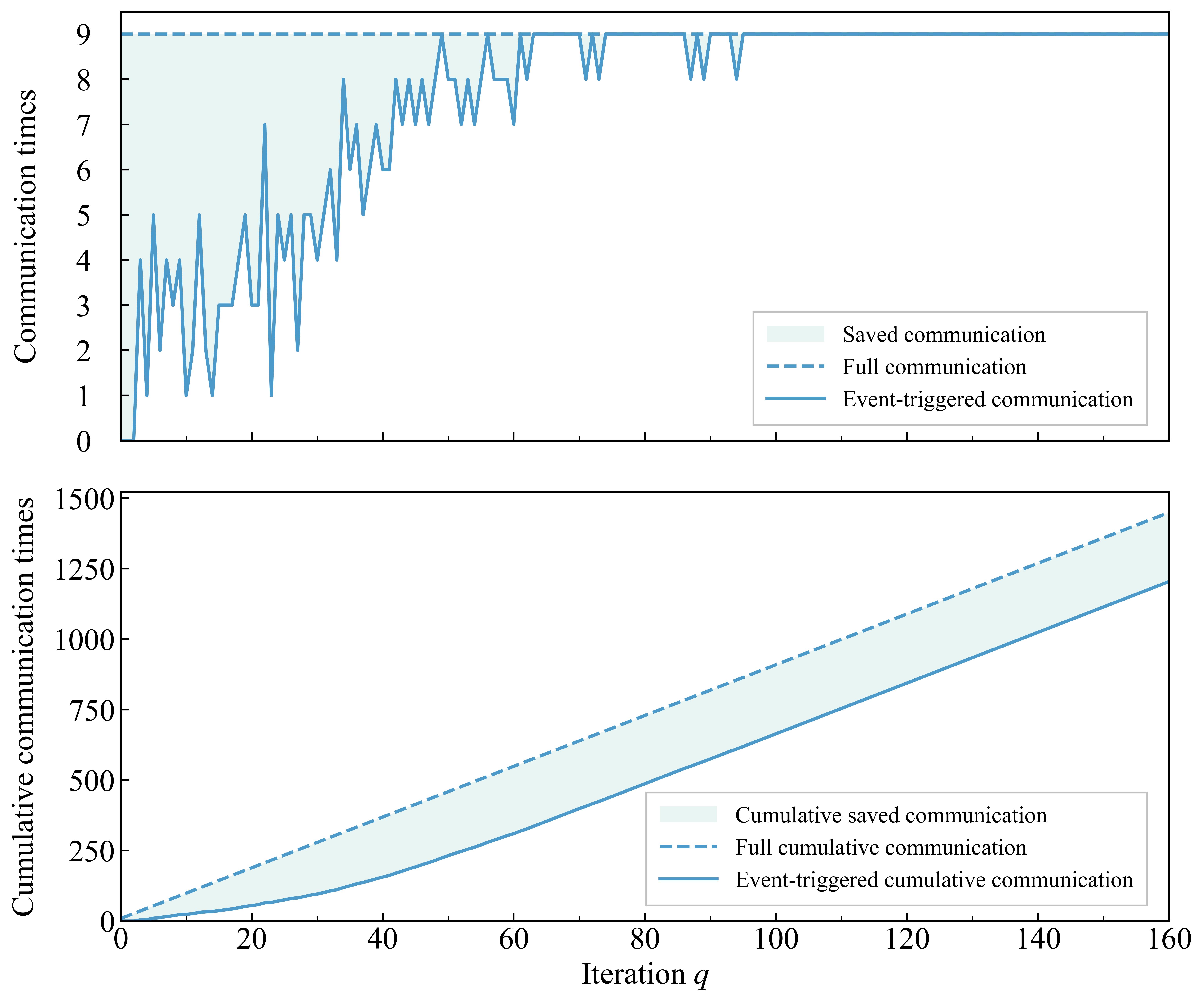}
	\caption{Communication performance of the event-triggered mechanism.}
	\label{fig:communication_performance}
\end{figure}
Figure~\ref{fig:state_control} shows the state and control-input trajectories in the upper and
lower subplots, respectively. The gray dashed lines denote the
corresponding constraints. It can be seen that all trajectories remain
within their prescribed bounds. 
The state components of all agents rapidly reach consensus, while the control inputs converge to steady values and remain nearly constant after consensus is reached.
Figure~\ref{fig:communication_performance} illustrates the communication performance of the event-triggered mechanism. The upper subplot shows the number of broadcast events occurring at each iteration~$q$, whereas the lower subplot depicts the cumulative number of broadcasts up to the corresponding iteration. 
Compared with broadcasting at every iteration, the proposed mechanism
reduces the overall communication load, particularly during the early iterations.

\subsection{Robust MPC-Based Consensus Control}

\begin{figure}[H]
	\centering
	\includegraphics[width=0.6\linewidth]{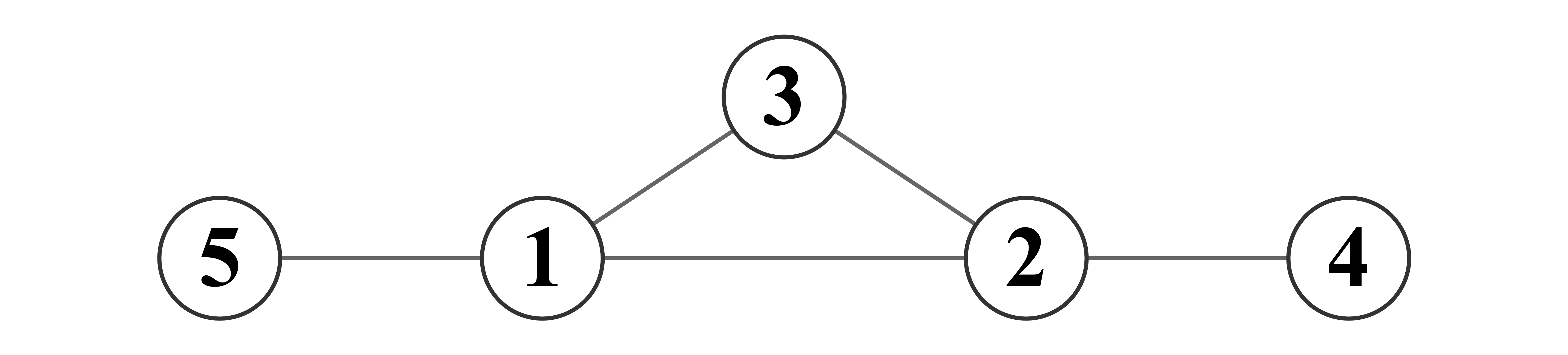}
	\caption{Communication topology of the five agents.}
	\label{fig:communication_topology_5}
\end{figure}

\begin{figure}[!htbp]
	\centering
	\includegraphics[width=0.9\linewidth]{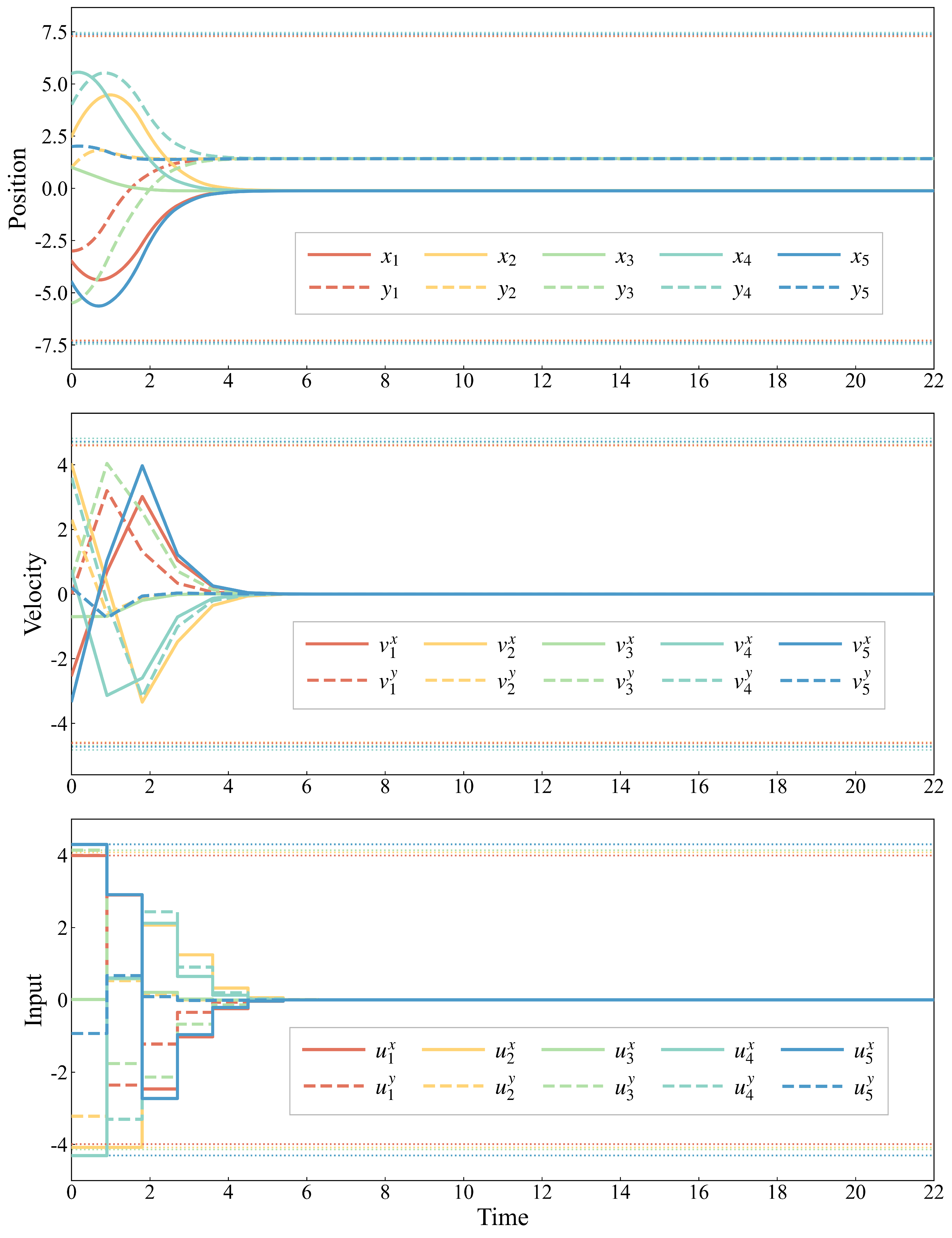}
	\caption{Closed-loop position, velocity, and control input trajectories for the system with five agents.}
	\label{fig:state_control_input}
\end{figure}

In this part, the proposed robust MPC-based consensus control
scheme is applied to a cooperative motion scenario involving five mobile
agents operating in a bounded two-dimensional workspace. The agents exchange
information over the communication topology shown in Figure~\ref{fig:communication_topology_5}.
Assume that
each agent is modeled as a two-dimensional second-order mechanical system
with heterogeneous mass $m_i$, described by
{\small
	\begin{align*}
		p_i(k+1)
		&=p_i(k)+\delta \dot{p}_i(k)
		+\tfrac{1}{2m_i}\delta^2 u_i(k),
		\nonumber\\
		\dot{p}_i(k+1)
		&=\dot{p}_i(k)+\tfrac{1}{m_i}\delta u_i(k).
	\end{align*}
}%
The common sampling interval $\delta$ of all agents is set to 
$0.9$ and is assumed to be known to the controller, while the heterogeneous 
masses $m_i$ are randomly selected from $[0.8,1.2]$ and assumed to be unknown 
to the controller a priori. Offline state-input 
data are collected under persistently exciting inputs in the presence of 
bounded measurement noise satisfying 
$\|d_i(k)\|\le \bar d_i=0.02$. The unknown
coefficient $1/m_i$ is then estimated from the noisy data using the least-squares 
procedure. 
Considering the limited operating region and actuator capability of the 
mobile agents, the admissible position, velocity, and control input sets are 
specified as 
$  
\mathbb{P}_i=  
\left\{  
p_i\in\mathbb{R}^{2}  
\,\middle|\,  
-8\le p_{i1},p_{i2}\le 8  
\right\}$,  
$  
\mathbb{V}_i=  
\left\{  
\dot{p}_i\in\mathbb{R}^{2}  
\,\middle|\,  
-5.5\le \dot{p}_{i1},\dot{p}_{i2}\le 5.5  
\right\}$,  
and  
$  
\mathbb{U}_i=  
\left\{  
u_i\in\mathbb{R}^{2}  
\,\middle|\,  
-5\le u_{i1},u_{i2}\le 5  
\right\}  
$.
The inner-loop distributed optimization in 
Algorithm~\ref{alg:dhgpadmm} is implemented with 
$P_{vi}=I_{16}$ and $p_{zi}=0.01$, 
$i\in\{1,2,\ldots,5\}$. The event-triggering threshold is chosen as 
$ 
\sigma_i^q=\sigma_i^0\beta^q, 
$ 
with $\sigma_i^0=5$ and $\beta=0.95$ for all agents. 
For the MPC problem, 
the weighting parameters are 
$q_{xi}=2$, $q_{vi}=1$, $r_i=0.1$, 
$p_{xi}=40$, and $p_{vi}=20$. 
The prediction horizon is $\mathcal T=8$, the control period is
$\Delta t=2$, and the closed-loop simulation is executed over
$T_S=25$ sampling steps. The agents operate with the common sampling 
interval, with each computed control input held constant over one sampling 
interval through a zero-order hold. 
Figure~\ref{fig:state_control_input} shows the closed-loop position, velocity,
and control input trajectories of the five agents. It can be observed that
the agents rapidly converge to the optimal consensus state, while all state
trajectories satisfy the prescribed constraints throughout the closed-loop
evolution. The control inputs also remain within their admissible bounds.
These results demonstrate the effectiveness of the proposed MPC framework.

\begin{figure}[!htbp]
	\centering
	\includegraphics[width=0.80\linewidth]{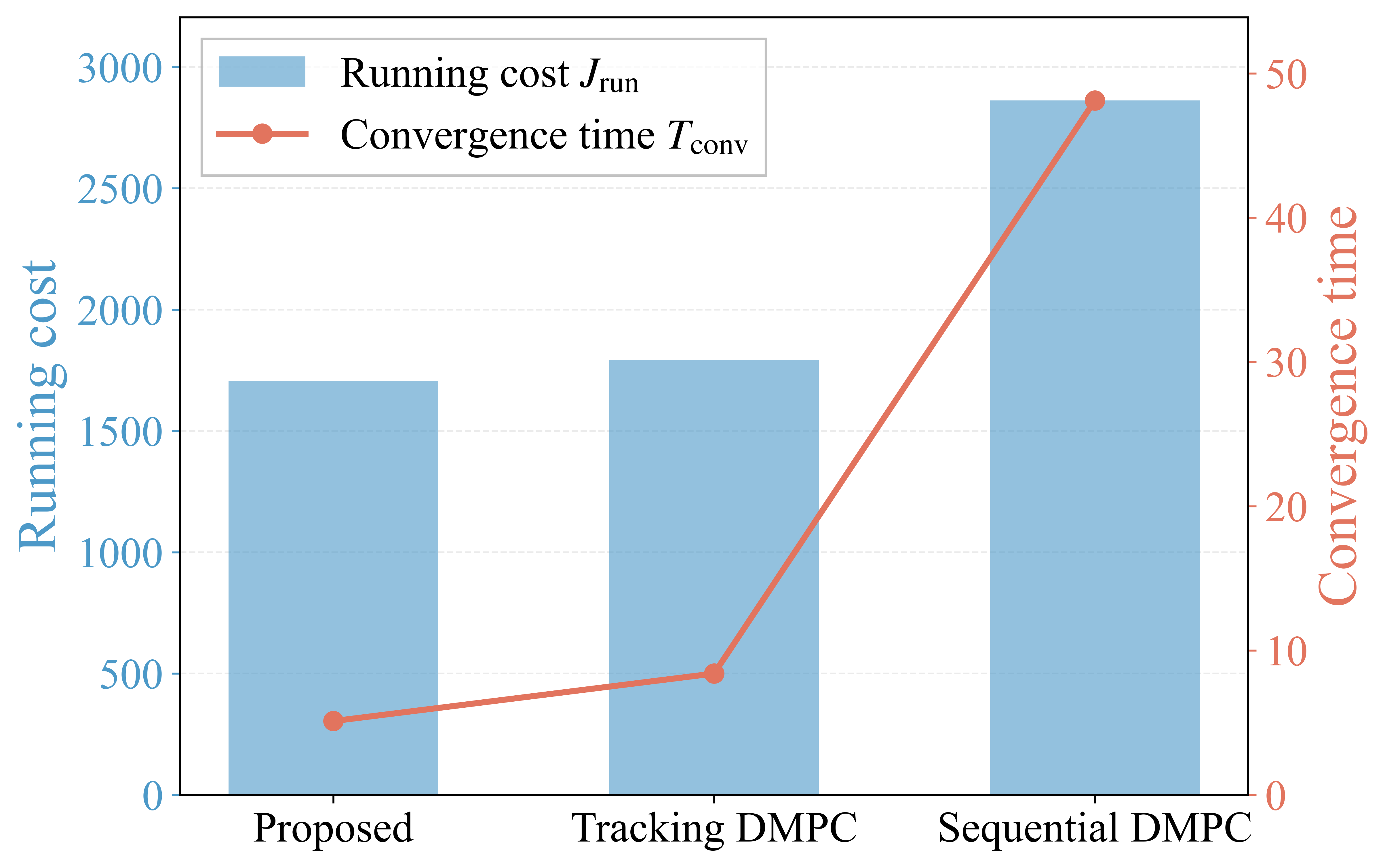}
	\caption{Comparison of the proposed method with the tracking DMPC and sequential DMPC methods.}
	\label{fig:comparison_three_methods}
\end{figure}
To further evaluate the proposed method, comparisons are conducted with 
a tracking DMPC method \cite{deng2025Distributed} and a sequential DMPC method 
\cite{hirche2020Distributed} in terms of the position convergence time defined by
$ 
T_{\mathrm{conv}} 
= 
\inf\{ 
t: 
\max_{i<j}\|p_i(\tau)-p_j(\tau)\|_2 
\leq \varepsilon_p,\; 
\forall \tau \geq t 
\}, 
\label{eq:conv_time_cmp} 
$ 
and the running cost defined by
$ 
J_{\mathrm{run}} 
= 
\sum_{k=0}^{k_{\mathrm{conv}}-1} 
(
\sum_{i=1}^{N}\sum_{j=1}^{N} 
a_{ij} q_{xi} 
\|p_i(k)-p_j(k)\|_2^2 + 
\sum_{i=1}^{N} 
r_i\|u_i(k)\|_2^2 
). 
$ 
Here, $k_{\mathrm{conv}}$ denotes the time step at which consensus is achieved, 
and $p_i(k)$ is the position of agent $i$. The same evaluation weights are used for all 
compared methods. The comparison results are shown in Figure~\ref{fig:comparison_three_methods}, 
which demonstrates that the proposed MPC framework achieves better consensus performance.

\section{Conclusion}
This article studies optimal consensus of heterogeneous multiagent systems with unknown dynamics. A robust MPC framework is developed to jointly optimize the control input and consensus sequences while enforcing state and input constraints, where noisy offline data are used to identify nominal agent models and quantify the associated model uncertainty. 
We design an event-triggered distributed gradient projection ADMM algorithm to solve the MPC optimization problem, and provide convergence conditions to ensure that the proposed algorithm converges to the optimal solution under event-triggered information exchange. Robust constraint tightening and invariant set arguments are further used to establish closed-loop guarantees. In particular, sufficient conditions are derived to guarantee robust constraint satisfaction and practical consensus of the closed-loop multiagent system under identification-induced model uncertainty.
Numerical
results validate that the proposed framework achieves robust consensus while reducing communication overhead and offers advantages in terms of closed-loop performance.

	\appendices
\section*{Appendix A -- Proof of Theorem \ref{theorem:1}}
Before analyzing the convergence of Algorithm 1, we first establish several key lemmas.
\begin{lemma}[{\cite{polyak1987introduction}}]
	On a closed convex set \(\mathbb{Q}\), the projection of \(x \in \mathbb{R}^n\) possesses $
	\langle x - \mathcal{P}_\mathbb{Q}(x), y - \mathcal{P}_\mathbb{Q}(x) \rangle \leq 0 ,~\forall x \in \mathbb{R}^n, y \in \mathbb{Q}$.
	\label{lemma:2form2024}
\end{lemma}
\begin{lemma}[\cite{drori2015Simple}]
	For any continuously differentiable convex function \(h\) on \(\mathbb{R}^p\) with \(L_h\)-Lipschitz continuous gradient \(\nabla h\), one has
	$	h(x) \leq h(y) + \langle \nabla h(z), x-y \rangle
	+ (L_h/2)\|x-z\|^2$, \(\forall x, y, z \in \mathbb{R}^p\).
	\label{lemma:3form2024}
\end{lemma}
\begin{lemma}[\cite{xu2017Accelerated}]
	Given a positive semidefinite matrix \(W \in \mathbb{R}^{n \times n}\), there has $
	2 \left\langle Wu, v \right\rangle = \left\| u \right\|_W^2 + \left\| v \right\|_W^2 - \left\| u - v \right\|_W^2,~\forall u, v \in \mathbb{R}^n$.\label{lemma:1from2021}
\end{lemma}
\begin{lemma}[\cite{xu2018Bregman}]
	If Assumption \ref{ass:graph} holds, then for each $Y \in (\operatorname{span}\{\mathbf{1}_N\} \otimes \mathbb{R}^{n(\mathcal T+1)})^\perp$, there exists a unique $Y' \in (\operatorname{span}\{\mathbf{1}_N\} \otimes \mathbb{R}^{n(\mathcal T+1)})^\perp$ such that $Y = \rho\tilde{\mathcal L}Y'$ and vice versa.
	\label{lemma:2form2021}
\end{lemma}

\textbf{Then, the proof of Theorem~\ref{theorem:1} is given below.}
\begin{proof}
Denote $P_V \triangleq \operatorname{diag}\{P_{v1}, P_{v2}, \ldots, P_{vN}\}$ and
$P_Z \triangleq \operatorname{diag}\{p_{z1}, p_{z2}, \ldots, p_{zN}\}\otimes I_{n(\mathcal T+1)}$ as the proximal matrices.
Then,
the proposed event-triggered algorithm (\ref{eq:update_u}), (\ref{eq:update_z}) and (\ref{eq:update_y}) can be expressed in the following compact matrix form
\begin{subequations}
	\begin{align}
		&\scalebox{0.9}{$\displaystyle
			V^{q+1}
			= \arg \min_{V \in \bar{\mathbb{V}}}
			\big\{ J(V, Z^q)
			+ \tfrac{1}{2} \| V - V^q \|_{P_V}^2 \big\}
			$}
		\label{eq:update_U}
		\\
		&\scalebox{0.9}{$\displaystyle
			\begin{aligned}
				Z^{q+1}
				&= \mathcal{P}_{\bar{\mathbb{Z}}}
				\Big\{ Z^q - P_Z
				\big[ \nabla_Z J(V^{q+1}, Z^q) 
				+ Y^q + \rho\tilde{\mathcal{L}}\hat{Z}^q
				\big] \Big\}
			\end{aligned}
			$}
		\label{eq:update_Z}
		\\
		&\scalebox{0.9}{$\displaystyle
			Y^{q+1}
			= Y^q + \rho\tilde{\mathcal{L}}\hat{Z}^{q+1}
			$}
		\label{eq:update_Y}
	\end{align}
\end{subequations}
where 
$\bar{\mathbb{V}} \triangleq \{ V=\operatorname{col}\{\mathbf{v}_1, \mathbf{v}_2, \dots, \mathbf{v}_N\} \in \mathbb{R}^{Nm\mathcal T} \mid \mathbf{v}_i \in \bar{\mathbb{V}}_i, \forall i \in \{1,\dots,N\} \bigr\},$ $	\bar{\mathbb{Z}} \triangleq \{ Z=\operatorname{col}\{\mathbf{z}_1, \mathbf{z}_2, \dots, \mathbf{z}_N\} \in \mathbb{R}^{Nn(\mathcal T+1)}\mid\mathbf{z}_i \in \bar{\mathbb{Z}}_i,\ \forall i \in \{1,\dots,N\}\bigr\}$.
From the optimal condition, the solution of (\ref{eq:update_U}) satisfies $\forall V \in \bar{\mathbb{V}}$
\begin{align}
	\scalebox{0.86}{$\displaystyle
		\left(V - V^{q+1}\right)^T
		\left[\nabla_V J(V^{q+1}, Z^q)
		+ P_V(V^{q+1} - V^q)\right]\ge 0.
		$}
	\label{eq:optimal_U}
\end{align}  
From Lemma \ref{lemma:2form2024} the solution of (\ref{eq:update_Z}) satisfies $\forall Z \in \bar{\mathbb{Z}}$
\begin{align}
	\scalebox{0.9}{$\displaystyle
		\begin{aligned}
			&\left(Z - Z^{q+1}\right)^T \big[ P_Z^{-1}(Z^{q+1} - Z^q) + \nabla_Z J(V^{q+1}, Z^q) \\
			&\qquad + Y^q + \rho\tilde{\mathcal{L}}\hat{Z}^q \big] \geq 0.
		\end{aligned}
		$}
	\label{eq:optimal_Z}
\end{align}
Adding (\ref{eq:optimal_U}) and (\ref{eq:optimal_Z}) and rearranging, we obtain
\begin{align}
	&\scalebox{0.9}{$\displaystyle
		(V - V^{q+1})^T \nabla_V J(V^{q+1}, Z^q)
		+ (Z - Z^{q+1})^T \nabla_Z J(V^{q+1}, Z^q)
		$} \nonumber\\
	&\scalebox{0.9}{$\displaystyle
		+ (V - V^{q+1})^T P_V (V^{q+1} - V^q)
		+ (Z - Z^{q+1})^T P_Z^{-1}(Z^{q+1} - Z^q)
		$} \nonumber\\
	&\scalebox{0.9}{$\displaystyle
		+ (Z - Z^{q+1})^T
		(Y^q + \rho\tilde{\mathcal{L}}\hat{Z}^q) \ge 0.
		$}
	\label{eq:UZ3parts}
\end{align}
%
We consider \eqref{eq:UZ3parts} in three parts. Applying Lemma~\ref{lemma:3form2024} agentwise to the first part yields
\begin{align} 
	&\scalebox{0.9}{$\displaystyle
		\left(V - V^{q+1}\right)^T \nabla_V J(V^{q+1}, Z^q)
		+ \left(Z - Z^{q+1}\right)^T \nabla_Z J(V^{q+1}, Z^q)
		$} \nonumber\\ 
	&\scalebox{0.9}{$\displaystyle
		= \left(V - V^{q+1}\right)^T \nabla_V J(V^{q+1}, Z^q)
		+ \left(Z - Z^{q}\right)^T \nabla_Z J(V^{q+1}, Z^q)
		$} \nonumber\\ 
	&\scalebox{0.9}{$\displaystyle
		\quad + \left(Z^{q} - Z^{q+1}\right)^T
		\nabla_Z J(V^{q+1}, Z^q)
		$} \nonumber\\ 
	&\scalebox{0.9}{$\displaystyle
		\le J(V,Z) - J(V^{q+1}, Z^{q+1})
		+ \tfrac{1}{2}\|Z^{q+1} - Z^q\|^2_{L_h},
		$}
	\label{eq:part1} 
\end{align}
where $L_h = \operatorname{diag}\{L_{h1}, L_{h2}, \dots, L_{hN}\} \otimes I_{n(\mathcal T+1)}$.
According to Lemma \ref{lemma:1from2021}, the second part can be expressed as
\begin{align*}
	\scalebox{0.9}{$\displaystyle
		\begin{aligned}
			&\big(V - V^{q+1}\big)^T P_V \big(V^{q+1} - V^q\big) + \big(Z - Z^{q+1}\big)^T P_Z^{-1} \big(Z^{q+1} - Z^q\big) \\
			&= \tfrac{1}{2}\big(-\left\| V-V^{q+1} \right\|_{P_V}^2 - \left\| V^{q+1}-V^{q} \right\|_{P_V}^2 + \left\| V-V^{q} \right\|_{P_V}^2\big) \\
			&\quad + \tfrac{1}{2}\big(\|Z - Z^q\|_{P_Z^{-1}}^2 - \|Z - Z^{q+1}\|_{P_Z^{-1}}^2 - \|Z^{q+1} - Z^q\|_{P_Z^{-1}}^2\big).
		\end{aligned}
		$}
\end{align*}
For any $Z \in \operatorname{null}(\tilde{\mathcal{L}})$ and $Y^{q+1} \in (\operatorname{span}\{\mathbf{1}_N\} \otimes \mathbb{R}^{n(\mathcal T+1)})^\perp$, we have $\langle Z, Y^{q+1} \rangle = 0$. Then, recalling $\hat{Z}^q = Z^q + e^q$, we obtain the third part
{\small	\begin{align*} 
		&\scalebox{0.9}{$\displaystyle 
			(Z - Z^{q+1})^T 
			[Y^q + \rho\tilde{\mathcal{L}}\hat{Z}^q] 
			$} \nonumber\\ 
		&\scalebox{0.9}{$\displaystyle 
			= \left\langle Z - Z^{q+1}, 
			Y^{q+1}+ \rho\tilde{\mathcal{L}} 
			(\hat{Z}^q-\hat{Z}^{q+1}) 
			\right\rangle 
			$} \nonumber\\ 
		&\scalebox{0.9}{$\displaystyle 
			= -\left\langle \hat{Z}^{q+1} - e^{q+1}, 
			Y^{q+1}-Y \right\rangle 
			- \langle Z^{q+1}, Y \rangle 
			$} \nonumber\\ 
		&\scalebox{0.9}{$\displaystyle 
			+ \left\langle Z - Z^{q+1}, 
			\rho\tilde{\mathcal{L}}({Z}^q-{Z}^{q+1}) 
			\right\rangle 
			+ \left\langle Z - Z^{q+1}, 
			\rho\tilde{\mathcal{L}}({e}^q-e^{q+1}) 
			\right\rangle 
			$} \nonumber\\ 
		&\scalebox{0.9}{$\displaystyle 
			= -\left\langle Y^{q+1}-Y^q, 
			Y'^{q+1}-Y' \right\rangle 
			+ \left\langle e^{q+1}, Y^{q+1}-Y \right\rangle 
			- \langle Z^{q+1}, Y \rangle 
			$} \nonumber\\ 
		&\scalebox{0.9}{$\displaystyle 
			\quad + \left\langle Z - Z^{q+1}, 
			\rho\tilde{\mathcal{L}}({Z}^q-{Z}^{q+1}) 
			\right\rangle 
			+ \left\langle Z - Z^{q+1}, 
			\rho\tilde{\mathcal{L}}({e}^q-e^{q+1}) 
			\right\rangle 
			$} \nonumber\\ 
		&\scalebox{0.9}{$\displaystyle 
			= -\tfrac{1}{2} \big( 
			\|Y'^q - Y'^{q+1}\|_{\rho\tilde{\mathcal{L}}}^2 
			+ \|Y' - Y'^{q+1}\|_{\rho\tilde{\mathcal{L}}}^2 
			- \|Y' - Y'^{q}\|_{\rho\tilde{\mathcal{L}}}^2 
			\big) 
			$} \nonumber\\ 
		&\scalebox{0.9}{$\displaystyle 
			\quad  
			+ \tfrac{1}{2} \big( 
			\|Z-Z^{q+1}\|_{\rho\tilde{\mathcal{L}}}^2 
			+ \|Z^q-Z^{q+1}\|_{\rho\tilde{\mathcal{L}}}^2 
			- \|Z-Z^q\|_{\rho\tilde{\mathcal{L}}}^2 
			\big) 
			$} \nonumber\\ 
		&\scalebox{0.9}{$\displaystyle 
			\quad + \left\langle e^{q+1}, 
			Y^{q+1}-Y \right\rangle 
			- \langle Z^{q+1}, Y \rangle 
			+ \left\langle Z - Z^{q+1}, 
			\rho\tilde{\mathcal{L}}({e}^q-e^{q+1}) 
			\right\rangle 
			$} \nonumber\\ 
		&\scalebox{0.9}{$\displaystyle 
			\stackrel{i}{=} 
			-\tfrac{1}{2} \big( 
			\|Y^q - Y^{q+1}\|_{M^{-1}}^2 
			+ \|Y - Y^{q+1}\|_{M^{-1}}^2 
			- \|Y - Y^{q}\|_{M^{-1}}^2 
			\big) 
			$} \nonumber\\ 
		&\scalebox{0.9}{$\displaystyle 
			\quad  
			+ \tfrac{1}{2} \big( 
			\|Z-Z^{q+1}\|_{\rho\tilde{\mathcal{L}}}^2 
			+ \|Z^q-Z^{q+1}\|_{\rho\tilde{\mathcal{L}}}^2 
			- \|Z-Z^q\|_{\rho\tilde{\mathcal{L}}}^2 
			\big) 
			$} \nonumber\\ 
		&\scalebox{0.9}{$\displaystyle 
			\quad + \left\langle e^{q+1}, 
			Y^{q+1}-Y \right\rangle 
			- \langle Z^{q+1}, Y \rangle 
			+ \left\langle Z - Z^{q+1}, 
			\rho\tilde{\mathcal{L}}({e}^q-e^{q+1}) 
			\right\rangle. 
			$}  
\end{align*}}%
Due to $Y, Y', Y'^q \in (\operatorname{span}\{\mathbf{1}_N\} \otimes \mathbb{R}^{n(\mathcal T+1)})^\perp$, we have $Y - Y^q = \rho\tilde{\mathcal{L}} (Y' - Y'^q)
= (
\rho\tilde{\mathcal{L}}
+ \frac{1}{N} \mathbf{1}\mathbf{1}^T
\otimes I_{n(\mathcal T+1)}
)
(Y' - Y'^q)$,
which further implies $Y' - Y'^q = M^{-1} (Y - Y^q)$, where $M = \rho\tilde{\mathcal{L}} + \frac{1}{N} \mathbf{1}_N\mathbf{1}_N^T \otimes I_{n(\mathcal T+1)}$. Then, we consider $
\|Y' - Y'^q\|_{\rho\tilde{\mathcal{L}}}^2 = \langle Y - Y^q, Y' - Y'^q \rangle = \|Y - Y^q\|_{M^{-1}}^2$. This identity justifies the equality marked by $\stackrel{i}{=}$.	
Then, by substituting the results of the three parts, we obtain the consolidated inequality
{
	\small
	\begin{align}
		&J(V,Z) - J(V^{q+1},Z^{q+1}) + \left\langle e^{q+1}, Y^{q+1}-Y \right\rangle - \langle Z^{q+1}, Y \rangle\nonumber \\
		&+ \tfrac{1}{2}\big(\| V-V^q \|_{P_V}^2 - \left\| V-V^{q+1} \right\|_{P_V}^2 - \left\| V^{q+1}-V^{q} \right\|_{P_V}^2 \big) \nonumber \\
		&+ \tfrac{1}{2}\big(\|Z - Z^q\|_{\Omega_Z}^2 - \|Z - Z^{q+1}\|_{\Omega_Z}^2 - \|Z^{q+1} - Z^q\|_{\Omega_Z - L_h}^2 \big) \nonumber \\
		&+ \tfrac{1}{2}\big( \|Y - Y^{q}\|_{M^{-1}}^2 - \|Y^q - Y^{q+1}\|_{M^{-1}}^2 - \|Y - Y^{q+1}\|_{M^{-1}}^2 \big) \nonumber \\
		&+ \left\langle Z - Z^{q+1}, \rho\tilde{\mathcal{L}}({e}^q-e^{q+1}) \right\rangle \ge 0,\nonumber
	\end{align}
}%
where $\Omega_Z \triangleq P_Z^{-1} - \rho\tilde{\mathcal{L}}$.
By the joint convexity of $J(V, Z)$ and KKT conditions, for any feasible $(V, Z)$, we have
{\small\begin{align}
		J(V, Z) - J(V^\star, Z^\star) &\ge \left\langle \nabla_V J(V^\star, Z^\star), V - V^\star \right\rangle \nonumber\\
		&\quad+ \left\langle \nabla_Z J(V^\star, Z^\star), Z - Z^\star \right\rangle \nonumber \\
		&= -\langle Y^\star, Z \rangle + \langle Y^\star, Z^\star \rangle 
		= -\langle Y^\star, Z \rangle, \nonumber
\end{align}}%
where $\langle Y^\star, Z^\star \rangle = 0$ holds due to $Z^\star \in \operatorname{null}(\tilde{\mathcal{L}})$ and $Y^\star \in (\operatorname{span}\{\mathbf{1}_N\} \otimes \mathbb{R}^{n(\mathcal T+1)})^\perp$. Rearranging the terms leads to
$
J(V, Z) - J(V^\star, Z^\star) + \langle Y^\star, Z \rangle \ge 0$, $\forall(V, Z)$.
Then, we let $V = V^\star$ and $Z = Z^\star$ and $Y = Y^\star$ and sum it over from $q = 0$ to $t - 1$, which yields
\begin{align} 
	&\scalebox{0.9}{$\displaystyle 
		0 \le \sum_{q=0}^{t-1} \big( 
		J(V^{q+1}, Z^{q+1}) - J(V^\star, Z^\star) 
		+ \langle Y^\star, Z^{q+1} \rangle 
		\big) 
		$} \nonumber\\ 
	&\scalebox{0.9}{$\displaystyle 
		\le \tfrac{1}{2}\big( 
		\| V^\star - V^0 \|_{P_V}^2 
		- \| V^\star - V^t \|_{P_V}^2 
		+ \| Z^\star - Z^0 \|_{\Omega_Z}^2 
		$} \nonumber\\ 
	&\scalebox{0.9}{$\displaystyle 
		\quad - \| Z^\star - Z^t \|_{\Omega_Z}^2 
		+ \| Y^\star - Y^0 \|_{M^{-1}}^2 
		- \| Y^\star - Y^t \|_{M^{-1}}^2 
		\big) 
		$} \nonumber\\ 
	&\scalebox{0.9}{$\displaystyle 
		\quad - \tfrac{1}{2}\Big( 
		\sum_{q=0}^{t-1} \| V^{q+1} - V^q \|_{P_V}^2 
		+ \sum_{q=0}^{t-1} \| Y^q - Y^{q+1} \|_{M^{-1}}^2 
		$} \nonumber\\ 
	&\scalebox{0.9}{$\displaystyle 
		\quad + \sum_{q=0}^{t-1} 
		\| Z^{q+1} - Z^q \|_{\Omega_Z - L_h}^2 
		\Big) 
		+ \sum_{q=0}^{t-1} 
		\langle e^{q+1}, Y^{q+1} - Y^\star \rangle 
		$} \nonumber\\ 
	&\scalebox{0.9}{$\displaystyle 
		\quad + \sum_{q=0}^{t-1} 
		\left\langle 
		Z^\star - Z^{q+1}, 
		\rho\tilde{\mathcal{L}}(e^q - e^{q+1}) 
		\right\rangle. 
		$} 
	\label{eq:telescoping_with_negatives_omega} 
\end{align}
Since $P_V \succ 0$, $M^{-1} \succ 0$, and $\Omega_Z - L_h \succ 0$, it holds
\begin{align}  
	&\scalebox{0.9}{$\displaystyle 
		\tfrac{1}{2}\big( 
		\| V^\star - V^t \|_{P_V}^2 
		+ \| Z^\star - Z^t \|_{\Omega_Z}^2 
		+ \| Y^\star - Y^t \|_{M^{-1}}^2 
		\big) 
		$} \nonumber \\  
	&\scalebox{0.9}{$\displaystyle 
		\le \tfrac{1}{2}\big( 
		\| V^\star - V^0 \|_{P_V}^2 
		+ \| Z^\star - Z^0 \|_{\Omega_Z}^2 
		+ \| Y^\star - Y^0 \|_{M^{-1}}^2 
		\big) 
		$} \nonumber \\  
	&\scalebox{0.9}{$\displaystyle 
		+ \sum_{k=1}^{t} \big( 
		\langle e^k, Y^k - Y^\star \rangle 
		+ \bigl\langle 
		Z^\star - Z^k, 
		\rho\tilde{\mathcal{L}}(e^{k-1} - e^k) 
		\bigr\rangle 
		\big). 
		$}\nonumber 
\end{align}%
Applying the Cauchy--Schwarz inequality together with the error threshold condition in Assumption~\ref{ass:threshold}, we obtain
\begin{align}  
	&\scalebox{0.9}{$\displaystyle  
		(\lambda_b/6)\left(  
		\| V^\star - V^t \|  
		+ \| Z^\star - Z^t \|  
		+ \| Y^\star - Y^t \|  
		\right)^2  
		$} 
	\label{eq:chained_four_parts_lambda_ab} \\  
	&\scalebox{0.9}{$\displaystyle 
		\le \tfrac{1}{2} \| V^\star - V^t \|_{P_V}^2 
		+ \tfrac{1}{2} \| Z^\star - Z^t \|_{\Omega_Z}^2 
		+ \tfrac{1}{2} \| Y^\star - Y^t \|_{M^{-1}}^2 
		$} \nonumber \\  
	&\scalebox{0.9}{$\displaystyle 
		\le \tfrac{1}{2} \| V^\star - V^0 \|_{P_V}^2 
		+ \tfrac{1}{2} \| Z^\star - Z^0 \|_{\Omega_Z}^2 
		+ \tfrac{1}{2} \| Y^\star - Y^0 \|_{M^{-1}}^2 
		$} \nonumber \\  
	&\scalebox{0.9}{$\displaystyle 
		\quad + \sum_{k=1}^{t} \big( 
		\langle e^k, Y^k - Y^\star \rangle 
		+ \bigl\langle 
		Z^\star - Z^k, 
		\rho\tilde{\mathcal{L}}(e^{k-1} - e^k) 
		\bigr\rangle 
		\big) 
		$} \nonumber \\  
	&\scalebox{0.9}{$\displaystyle 
		\le \tfrac{1}{2} \| V^\star - V^0 \|_{P_V}^2 
		+ \tfrac{1}{2} \| Z^\star - Z^0 \|_{\Omega_Z}^2 
		+ \tfrac{1}{2} \| Y^\star - Y^0 \|_{M^{-1}}^2 
		$} \nonumber \\  
	&\scalebox{0.9}{$\displaystyle 
		+ \lambda_a \sqrt{N} \sum_{k=1}^{t} \sigma^{k-1} 
		\big( 
		\|V^\star - V^k\| 
		+ \|Z^\star - Z^k\| 
		+ \|Y^\star - Y^k\| 
		\big), 
		$} 
	\nonumber 
\end{align}
where $\lambda_b \triangleq \min \{ \underline{\lambda}(P_V), \underline{\lambda}(\Omega_Z), {1}/{\overline{\lambda}(M)} \}$ and $\lambda_a \triangleq \max\{1, 2\rho\overline{\lambda}(\tilde{\mathcal{L}})\}$ are scale coefficients.
Using \cite[Lemma~1]{schmidt2011convergence}, we obtain $
\| V^\star - V^t \| 
+ \| Z^\star - Z^t \| 
+ \| Y^\star - Y^t \|
\le [
\frac{3}{\lambda_b}(
\| V^\star - V^0 \|_{P_V}^2
+ \| Z^\star - Z^0 \|_{\Omega_Z}^2 
+ \| Y^\star - Y^0 \|_{M^{-1}}^2
) + A_t^2
]^{1/2} + A_t
$,
where $A_t \triangleq {3\lambda_a\sqrt{N}}/{\lambda_b} \sum_{k=1}^{t} \sigma^{k-1}$. Since $\sigma^k \ge 0$, $A_t$ is non-decreasing, implying $A_k \le A_t$ for all $k \le t$. Therefore, the following uniform bound holds true for all $k \le t$
\begin{align}  
	&\scalebox{0.9}{$\displaystyle
		\| V^\star - V^k \|
		+ \| Z^\star - Z^k \|
		+ \| Y^\star - Y^k \|
		$} \nonumber \\  
	&\scalebox{0.9}{$\displaystyle
		\le 2A_t + \sqrt{3/\lambda_b}\,\mathcal{D}_0
		\triangleq M_t,
		$}
	\label{eq:residual_bound_uniform}  
\end{align}
where $\mathcal{D}_0 \triangleq \|V^\star - V^0\|_{P_V} + \|Z^\star - Z^0\|_{\Omega_Z} + \|Y^\star - Y^0\|_{M^{-1}}$. Note that under Assumption~\ref{ass:threshold}, $A_t$ is uniformly bounded, thereby ensuring that $M_t$ is uniformly bounded for all $t \ge 0$.
By rearranging and relaxing the terms in \eqref{eq:telescoping_with_negatives_omega}, and further bounding the cross-product terms via the Cauchy–Schwarz inequality and the event-triggering mechanism governed by Assumption~\ref{ass:threshold} while incorporating \eqref{eq:residual_bound_uniform}, we obtain 
\begin{align}
	& \scalebox{0.8}{$\displaystyle
		\frac{1}{2} \sum_{q=0}^{t-1} \Big(
		\|V^{q+1} - V^q\|_{P_V}^2
		+ \|Z^{q+1} - Z^q\|_{\Omega_Z - L_h}^2
		+ \|Y^q - Y^{q+1}\|_{M^{-1}}^2
		\Big)
		$} \nonumber \\
	& \scalebox{0.8}{$\displaystyle
		\le \tfrac{1}{2} \|V^\star - V^0\|_{P_V}^2
		+ \tfrac{1}{2} \|Z^\star - Z^0\|_{\Omega_Z}^2
		+ \tfrac{1}{2} \|Y^\star - Y^0\|_{M^{-1}}^2
		$} \nonumber \\
	& \scalebox{0.8}{$\displaystyle
		~~~+ \sum_{q=1}^{t} \langle e^q, Y^q - Y^\star \rangle
		+ \sum_{q=1}^{t} \langle Z^\star - Z^q,
		\rho\tilde{\mathcal{L}}(e^{q-1} - e^q) \rangle
		$} \nonumber \\
	& \scalebox{0.8}{$\displaystyle
		\le \tfrac{1}{2} \|V^\star - V^0\|_{P_V}^2
		+ \tfrac{1}{2} \|Z^\star - Z^0\|_{\Omega_Z}^2
		+ \tfrac{1}{2} \|Y^\star - Y^0\|_{M^{-1}}^2
		$} \nonumber \\
	& \scalebox{0.8}{$\displaystyle
		~~~ + \lambda_a M_t \sqrt{N} \sum_{q=1}^{t} \sigma^{q-1}.
		$}
	\label{eq:chained_partial_sum_bound}
\end{align}
Since the upper bound in \eqref{eq:chained_partial_sum_bound} is finite, and each summand on the left is nonnegative, the series converges. Under the parameter conditions of Theorem~\ref{theorem:1}, it follows that
$\lim_{q\to\infty} \|V^{q+1} - V^q\| = 0$,
$\lim_{q\to\infty} \|Z^{q+1} - Z^q\| = 0$,
$\lim_{q\to\infty} \|Y^{q+1} - Y^q\| = 0$.
Moreover, Assumption~\ref{ass:threshold} implies
$\sigma^q\rightarrow0$ as $q\rightarrow\infty$.
Since the communication error satisfies
\(
\|e^q\|
\leq
\sqrt{N}\sigma^q,
\)
we have
\(
e^q\rightarrow0.
\)
By \eqref{eq:residual_bound_uniform}, together with the uniform boundedness of
$A_t$ under Assumption~\ref{ass:threshold}, the sequence
$\{(V^q,Z^q,Y^q)\}$ is bounded.
Hence, it admits at least one cluster point.
Let
\(
(V^\infty,Z^\infty,Y^\infty)
\)
be an arbitrary cluster point, and let
$\{q_l\}$ be a subsequence such that
\(
(V^{q_l},Z^{q_l},Y^{q_l})
\rightarrow
(V^\infty,Z^\infty,Y^\infty).
\)
Since
\(
\|V^{q+1}-V^q\|\rightarrow0,~
\|Z^{q+1}-Z^q\|\rightarrow0,~
\|Y^{q+1}-Y^q\|\rightarrow0,
\)
we have
\(
(V^{q_l+1},Z^{q_l+1},Y^{q_l+1})
\rightarrow
(V^\infty,Z^\infty,Y^\infty).
\)
Furthermore, since $e^{q_l}\rightarrow0$ and
$e^{q_l+1}\rightarrow0$, we have
\(
\hat Z^{q_l}\rightarrow Z^\infty
\)
and
\(
\hat Z^{q_l+1}\rightarrow Z^\infty.
\)
From the dual update
\(
Y^{q+1}-Y^q
=
\rho\tilde{\mathcal L}\hat Z^{q+1},
\)
together with
$\|Y^{q+1}-Y^q\|\rightarrow0$ and
$\hat Z^{q_l+1}\rightarrow Z^\infty$, we obtain
\begin{align}
	&\scalebox{0.9}{$\displaystyle
		\tilde{\mathcal L}Z^\infty=\mathbf 0
		$}
	\label{eq:kkt_limit_consensus}
\end{align}
Taking $\ell\rightarrow\infty$ in \eqref{eq:optimal_U} and
\eqref{eq:optimal_Z}, and using
\eqref{eq:kkt_limit_consensus}, yields
\begin{align} 
	&\scalebox{0.9}{$\displaystyle
		(V-V^\infty)^T 
		\nabla_V J(V^\infty,Z^\infty) 
		\geq 0,
		\quad \forall V\in\bar{\mathbb V}
		$}
	\label{eq:kkt_limit_v}\\
	&\scalebox{0.9}{$\displaystyle
		(Z-Z^\infty)^T 
		\bigl(
		\nabla_Z J(V^\infty,Z^\infty)+Y^\infty 
		\bigr) 
		\geq 0,
		\quad \forall Z\in\bar{\mathbb Z}
		$}
	\label{eq:kkt_limit_z} 
\end{align}
Since $Y^0=\mathbf 0$ and every dual update belongs to
$\operatorname{range}(\tilde{\mathcal L})$, it follows that
$
Y^\infty\in\operatorname{range}(\tilde{\mathcal L}).
$
Therefore,
\eqref{eq:kkt_limit_consensus}--\eqref{eq:kkt_limit_z}
constitute the KKT conditions of
Problem~\ref{prob:optimal_consensus}.
Since Problem~\ref{prob:optimal_consensus} is convex and satisfies
$
\operatorname{ri}(\bar{\mathbb Z})
\cap
\ker(\tilde{\mathcal L})
\neq\emptyset,
$
the KKT conditions are necessary and sufficient for optimality.
Hence,
\(
(V^\infty,Z^\infty)
\)
is an optimal solution $(V^\star,Z^\star)$ of
Problem~\ref{prob:optimal_consensus}.
Since $J(V,Z)$ is jointly strictly convex, the primal optimal solution
$(V^\star,Z^\star)$ is unique.
As $(V^\infty,Z^\infty)$ is chosen as an arbitrary primal cluster point,
every cluster point of $\{(V^q,Z^q)\}$ must coincide
with $(V^\star,Z^\star)$. Moreover,
since the primal sequence is bounded and has a unique cluster point, we have
\(\lim_{q\to\infty}
(V^q,Z^q)=(V^\star,Z^\star)
\),
which completes the proof.
\end{proof} 

\section*{Appendix B -- Proof of Theorem \ref{theorem:2}}
\begin{proof}
Reorganizing \eqref{eq:telescoping_with_negatives_omega}, and combining with \eqref{eq:chained_four_parts_lambda_ab}--\eqref{eq:residual_bound_uniform}, we obtain
\begin{align}  
	\scalebox{0.9}{$\displaystyle 0$} 
	&\scalebox{0.9}{$\displaystyle 
		\le \sum_{q=0}^{t-1} \big( 
		J(V^{q+1}, Z^{q+1}) - J(V^\star, Z^\star) 
		+ \langle Y^\star, Z^{q+1} \rangle 
		\big) 
		$} \nonumber\\  
	&\scalebox{0.9}{$\displaystyle 
		\le \tfrac{1}{2}\mathcal{S}_0 
		+ \sum_{q=0}^{t-1} \big( 
		\langle e^{q+1}, Y^{q+1} - Y^\star \rangle 
		+ \langle Z^\star - Z^{q+1}, 
		\rho\tilde{\mathcal L}(e^q - e^{q+1}) \rangle 
		\big) 
		$} \nonumber\\  
	&\scalebox{0.9}{$\displaystyle 
		\le \tfrac{1}{2} \mathcal{S}_0 
		+ \lambda_a \sqrt{N} 
		\sum_{k=1}^{t} \sigma^{k-1} \mathcal{D}_k 
		\le \big( 
		\tfrac{\mathcal{D}_0}{\sqrt{2}} 
		+ \sqrt{\tfrac{2\lambda_b}{3}} A_t 
		\big)^2, 
		$} 
	\label{eq:theorem2_19}  
\end{align}
where $\mathcal{S}_0 \triangleq \|V^\star - V^0\|_{P_V}^2 + \|Z^\star - Z^0\|_{\Omega_Z}^2 + \|Y^\star - Y^0\|_{M^{-1}}^2$ and $\mathcal{D}_k \triangleq \|V^\star - V^k\| + \|Z^\star - Z^k\| + \|Y^\star - Y^k\|$.
By the convexity of $J$, we have
\begin{align}  
	&\scalebox{0.9}{$\displaystyle 
		J(\bar{V}_t, \bar{Z}_t) - J(V^\star, Z^\star) 
		+ \langle Y^\star, \bar{Z}_t \rangle 
		$} \nonumber \\  
	&\scalebox{0.9}{$\displaystyle 
		\le \frac{1}{t} \sum_{q=0}^{t-1} \big( 
		J(V^{q+1}, Z^{q+1}) - J(V^\star, Z^\star) 
		+ \langle Y^\star, Z^{q+1} \rangle 
		\big) 
		$} \nonumber \\  
	&\scalebox{0.9}{$\displaystyle  
		\le \tfrac{1}{2t} \big(  
		\mathcal{D}_0 + 2\sqrt{\lambda_b/3}\, A_t  
		\big)^2.  
		$}  
	\label{eq:theorem2_23}  
\end{align}
Since $Y^\star = \tilde{\mathcal L}\Lambda^\star$ and $Y^0 =\mathbf  0$, we have
$
\mathcal{D}_0
\le \|V^\star - V^0\|_{P_V} + \|Z^\star - Z^0\|_{\Omega_Z} + c\|\Lambda^\star\|
$,
where $c \triangleq \|\tilde{\mathcal L}\| \cdot \|M^{-1}\|^{1/2}$.
Taking the supremum over all $\Lambda^\star$ with $\|\Lambda^\star\| \le \gamma$ in \eqref{eq:theorem2_23} yields
\begin{align} 
	&\scalebox{0.9}{$\displaystyle 
		\sup_{\|\Lambda^\star\| \le \gamma} 
		\Big( 
		J(\bar{V}_t, \bar{Z}_t) - J(V^\star, Z^\star) 
		+ \langle \Lambda^\star, \tilde{\mathcal L}\bar{Z}_t\rangle 
		\Big) 
		$} \nonumber \\ 
	&\scalebox{0.9}{$\displaystyle 
		= J(\bar{V}_t, \bar{Z}_t) - J(V^\star, Z^\star) 
		+ \gamma \|\tilde{\mathcal L}\bar{Z}_t\| 
		\le (2t)^{-1}\Gamma_t^2,
		$} 
	\label{eq:theorem2_sup} 
\end{align}
where $\Gamma_t = \|V^\star - V^0\|_{P_V} + \|Z^\star - Z^0\|_{\Omega_Z} + c\gamma + \frac{2\sqrt{\lambda_b}}{\sqrt{3}} A_t$.
By the convexity of $J$ and the KKT conditions, for any $(V,Z)$, we have
$
J(V,Z) - J(V^\star, Z^\star) \ge -\langle Y^\star, Z - Z^\star\rangle$.
Taking $V = \bar{V}_t$ and $Z = \bar{Z}_t$, and noting that $\langle Y^\star, Z^\star\rangle = 0$, we have
\begin{align} 
	&\scalebox{0.9}{$\displaystyle
		J(\bar{V}_t, \bar{Z}_t) - J(V^\star, Z^\star) 
		\ge -\langle Y^\star, \bar{Z}_t\rangle 
		\ge -\|\Lambda^\star\| {\scriptstyle\times} \|\tilde{\mathcal L}\bar{Z}_t\|.
		$}
	\label{eq:lowerbound_jandj} 
\end{align}
From \eqref{eq:theorem2_sup} and \eqref{eq:lowerbound_jandj}, it follows that the consensus error satisfies
\begin{align}  
	&\scalebox{0.9}{$\displaystyle 
		\|\tilde{\mathcal L}\bar{Z}_t\|  
		\le 
		\bigl[2t(\gamma-\|\Lambda^\star\|)\bigr]^{-1}\Gamma_t^2.
		$} 
	\label{eq:theorem2_consensus}  
\end{align}
Substituting \eqref{eq:theorem2_consensus} into \eqref{eq:lowerbound_jandj} yields the lower bound for the objective error
\begin{align} 
	&\scalebox{0.9}{$\displaystyle
		-\tfrac{\|\Lambda^\star\|}{2t(\gamma - \|\Lambda^\star\|)} \Gamma_t^2 
		\le J(\bar{V}_t, \bar{Z}_t) - J(V^\star, Z^\star).
		$}
	\label{eq:theorem2_lower} 
\end{align}
Combining \eqref{eq:theorem2_sup} and \eqref{eq:theorem2_lower} we obtain the objective error bounds $
-\frac{\|\Lambda^\star\|}{2t(\gamma - \|\Lambda^\star\|)} \Gamma_t^2
\le J(\bar{V}_t, \bar{Z}_t) - J(V^\star, Z^\star)
\le \frac{1}{2t} \Gamma_t^2$.
Since $A_t$ is uniformly bounded, $\Gamma_t$ is also uniformly bounded. Therefore, both the objective error and the consensus error converge to zero at a rate of $O(1/t)$. This completes the proof of Theorem~\ref{theorem:2}.
\end{proof}

\section*{Appendix C -- Proof of Lemma \ref{lem:invariant_ellipsoid}}
\begin{proof}
Fix an arbitrary agent $i\in\{1,\dots,N\}$. For notational simplicity, the subscript $i$ is omitted throughout this proof.
Consider the quadratic function $V(\tilde e) = \tilde e^T P \tilde e$ with $P \succ 0$. Along the error dynamics \eqref{eq:error_dynamics}, $V(\tilde e(k+1))$ can be expanded as
$
V(\tilde e(k+1)) = \tilde e(k)^{T} A_K^{T} P A_K \tilde e(k) + 2\tilde e(k)^{T} A_K^{T} P \omega(k) + \omega(k)^{T} P \omega(k)
$.
By applying the condition \eqref{eq:condition_p} and invoking Young's
inequality with a scalar parameter $\vartheta\in(0,1)$ to bound the
cross-term, the difference
$V(\tilde e(k+1))-V(\tilde e(k))$ can be bounded as
\begin{align} 
	&\scalebox{0.95}{$\displaystyle
		V(\tilde e(k+1))-V(\tilde e(k))
		$} 
	\nonumber\\ 
	&\scalebox{0.95}{$\displaystyle
		\quad\le 
		-(1-\vartheta)\tilde e(k)^TW\tilde e(k) 
		+\omega(k)^T 
		\left(P+\Pi/\vartheta\right)\omega(k)
		$} 
	\nonumber\\ 
	&\scalebox{0.95}{$\displaystyle
		\quad\le 
		-c_e\|\tilde e(k)\|^2+d_\omega\bar\omega^2,
		$}
	\label{eq:difference_v} 
\end{align}
where
$c_e=(1-\vartheta)\underline{\lambda}(W)$,
$d_\omega=\bar{\lambda}\big(P+\vartheta^{-1}\Pi\big)$, and
$\Pi=PA_KW^{-1}A_K^TP$.
To establish robust forward invariance of $\mathcal{E}$, we must guarantee
$V(\tilde e(k+1)) \le \alpha$ for all $\tilde e(k) \in \mathcal{E}$.
Since
$V(\tilde e(k))=\tilde e(k)^T P\tilde e(k)
\le \bar{\lambda}(P)\|\tilde e(k)\|^2$,
it follows from \eqref{eq:difference_v} that $
V(\tilde e(k+1)) 
\le 
\left(1-c_e/\bar{\lambda}(P)\right)V(\tilde e(k)) 
+d_\omega\bar{\omega}^2$.
From \eqref{eq:condition_p}, we have
$W\prec P-A_K^TPA_K\preceq P$. Therefore, since
$W\succ0$ and $0<\vartheta<1$,
$0<c_e=(1-\vartheta)\underline{\lambda}(W)<\bar{\lambda}(P)$.
Hence, using $V(\tilde e(k))\le\alpha$, we have $
V(\tilde e(k+1)) 
\le 
\left(1-c_e/\bar{\lambda}(P)\right)\alpha 
+d_\omega\bar{\omega}^2$.
Moreover, by \eqref{eq:condition_alpha}, we have
$
d_\omega\bar{\omega}^2
\le
\frac{c_e}{\bar{\lambda}(P)}\alpha.
$
Therefore, $V(\tilde e(k))\le\alpha$ implies
$V(\tilde e(k+1))\le\alpha$ for all
$\|\omega(k)\|\le\bar{\omega}$.
Hence, $\mathcal{E}$ is robustly forward invariant. Since agent $i$ was chosen arbitrarily, the result holds for all agents.
\end{proof}

\section*{Appendix D -- Proof of Theorem \ref{theorem:first_order_closed_loop}}
\begin{proof}
Assume that $v_i^\star(\tau|t_k)$ and $z_i^\star(\tau|t_k)$ are the optimal solutions at the predictive time $t_k$ with $\tau \in [t_k, t_k+\mathcal{T}]$. A feasible solution at $t_{k+1}$ can be constructed as
\begin{align} 
	&\scalebox{0.9}{$\displaystyle
		\bar v_i(\tau|t_{k+1}) = 
		\begin{cases} 
			v_i^\star(\tau|t_k), \ \qquad \quad \tau \in [t_{k+1}, t_k+\mathcal{T}-1], \\ 
			\big[ 
			\bar z_i(\tau|t_{k+1}) 
			- 
			\bar{\hat x}_i(\tau|t_{k+1}) 
			\big] /\hat\kappa_i, \\ 
			\qquad \qquad \qquad \quad\tau \in [t_k+\mathcal{T}, t_{k+1}+\mathcal{T}-1], 
		\end{cases}
		$}
	\label{eq:tk1v} \\ 
	&\scalebox{0.9}{$\displaystyle
		\bar z_i(\tau|t_{k+1}) = 
		\begin{cases} 
			z_i^\star(\tau|t_k), & \tau \in [t_{k+1}, t_k+\mathcal{T}], \\ 
			z_i^\star(t_k+\mathcal{T}|t_k), & \tau \in [t_k+\mathcal{T}+1, t_{k+1}+\mathcal{T}], 
		\end{cases}
		$}
	\label{eq:tk1z} \\ 
	&\scalebox{0.9}{$\displaystyle
		\bar{\hat x}_i(\tau|t_{k+1}) = 
		\begin{cases} 
			\hat x_i^\star(\tau|t_k), & \tau \in [t_{k+1}, t_k+\mathcal{T}], \\ 
			z_i^\star(t_k+\mathcal{T}|t_k), & \tau \in [t_k+\mathcal{T}+1, t_{k+1}+\mathcal{T}]. 
		\end{cases}
		$}
	\label{eq:tk1x} 
\end{align}
Under condition \eqref{eq:first_order_feasibility_condition}, the constructed solution satisfies all constraints in Problem~\ref{prob:general_robust_mpc}, hence it is feasible.
For the actual system, since $\tilde e_i(t_0)\in\mathcal E_i$,
Lemma~\ref{lem:invariant_ellipsoid} guarantees
$\tilde e_i(k)\in\mathcal E_i$ for all $k\geq t_0$.
With $\hat x_i(k) \in \mathbb{X}_i \ominus \mathcal{E}_i$ and $v_i(k) \in \mathbb{U}_i \ominus K_i\mathcal{E}_i$, we have $x_i(k) = \hat x_i(k) + \tilde e_i(k) \in (\mathbb{X}_i \ominus \mathcal{E}_i) \oplus \mathcal{E}_i \subseteq \mathbb{X}_i$ and
$u_i(k) = v_i(k) + K_i \tilde e_i(k) \in (\mathbb{U}_i \ominus K_i\mathcal{E}_i) \oplus K_i\mathcal{E}_i \subseteq \mathbb{U}_i$.
Hence, the original constraints are always satisfied.
Consider the Lyapunov function $W(t_k) = J_{t_k}\big(V^\star(\tau|t_k),$ $Z^\star(\tau|t_k)\big) + \tilde e(t_k)^T P \tilde e(t_k).$ For the optimal control input $v^\star_i(\tau|t_{k+1})$ and $z^\star_i(\tau|t_{k+1})$ at time $t_{k+1}$, we have
{\small	\begin{align*}
		&W(t_{k+1}) - W(t_k) \\
		& \le \big[J_{t_{k+1}}\big(\bar V(\tau|t_{k+1}), \bar Z(\tau|t_{k+1})\big) - J_{t_k}\big(V^\star(\tau|t_k), Z^\star(\tau|t_k)\big)\big] \\
		&\quad - c_e\|\tilde e(t_k)\|^2 + d_\omega\|\omega(t_k)\|^2 \\
		& = \sum_{i=1}^{N} \Big\{
		\sum_{\tau=t_{k+1}}^{t_{k+1}+\mathcal{T}-1}
		\Big[ q_{si} \|\bar{\hat x}_i(\tau|t_{k+1}) - \bar z_i(\tau|t_{k+1})\|^2 \\
		&\quad + r_{si} \|\bar v_i(\tau|t_{k+1})\|^2 \Big] + p_{si} \|\bar{\hat x}_i(t_{k+1}+\mathcal{T}|t_{k+1})  \\
		&\quad - \bar z_i(t_{k+1}+\mathcal{T}|t_{k+1})\|^2 \Big\} - c_e\|\tilde e(t_k)\|^2 + d_\omega\|\omega(t_k)\|^2\\
		&\quad - \sum_{i=1}^{N} \Big\{
		\sum_{\tau=t_k}^{t_k+\mathcal{T}-1}
		\big[ q_{si} \|\hat x_i^\star(\tau|t_k) - z_i^\star(\tau|t_k)\|^2 + r_{si} \|v_i^\star(\tau|t_k)\|^2 \big] \\
		&\quad + p_{si} \|\hat x_i^\star(t_k+\mathcal{T}|t_k) - z_i^\star(t_k+\mathcal{T}|t_k)\|^2 \Big\}.
\end{align*}}%
Substituting the feasible solution in \eqref{eq:tk1v}--\eqref{eq:tk1x} yields
{\small	\begin{align*}
		&W(t_{k+1}) - W(t_k) \\
		&\quad \le \sum_{i=1}^N \Big\{
		( q_{si} + \tfrac{r_{si}}{\hat\kappa_i^2} - p_{si} ) \|\hat x_i^\star(t_k+\mathcal{T}|t_k) - z_i^\star(t_k+\mathcal{T}|t_k)\|^2 \\
		&\qquad - \sum_{\tau=t_k}^{t_{k+1}-1} \big[ q_{si}\|\hat x_i^\star(\tau|t_k) - z_i^\star(\tau|t_k)\|^2 + r_{si}\|v_i^\star(\tau|t_k)\|^2 \big] \Big\} \\
		&\quad\quad - c_e\|\tilde e(t_k)\|^2 + d_\omega\|\omega(t_k)\|^2.
\end{align*}}%
Under \eqref{eq:first_order_stability_condition}, the first term is non-positive.
Dropping nonnegative terms and noting that the double sum contains
$\|\hat x_i(t_k)-z_i^\star(t_k|t_k)\|^2$ as its first component, we have $W(t_{k+1}) - W(t_k)\le
-\alpha_x\|\hat x(t_k)-z^\star(t_k)\|^2
-c_e\|\tilde e(t_k)\|^2
+d_\omega\|\omega(t_k)\|^2$,
where $\alpha_x = \min_i\{q_{si}\} > 0$.
Since the Lyapunov function $W(t_k)$ satisfies the ISS-Lyapunov inequality,
it follows from ISS theory \cite{marruedo2002Inputtostate}
that the augmented system with state
$[(\hat x(t_k)-z^\star(t_k))^T,\tilde e(t_k)^T]^T$
is ISS. Since
$x(t_k)-z^\star(t_k)
=
[I\;I]
[(\hat x(t_k)-z^\star(t_k))^T,\tilde e(t_k)^T]^T$, there exist $\beta\in\mathcal{KL}$ and $r\in\mathcal K$ such that $
\|x(t_k)-z^\star(t_k)\|
\leq
\beta\bigl(
\|(\hat x(t_0)-z^\star(t_0),\tilde e(t_0))\|,
k
\bigr)
+r(\bar\omega)$.
This inequality implies that the tracking error
$\|x(t_k)-z^\star(t_k)\|$ is ultimately bounded by
$r(\bar\omega)$. 
Since $\Delta t=1$, the MPC problem is updated at every time
instant after $t_0$.
Moreover, \eqref{LZ} gives
$z_i^\star(k|k)=z_j^\star(k|k)$ for all $i,j$. Hence, we have $
\|x_i(k)-x_j(k)\|
\leq
\|x_i(k)-z_i^\star(k|k)\|
+
\|x_j(k)-z_j^\star(k|k)\|$,
which completes the proof.
\end{proof}

\section*{Appendix E -- Proof of Theorem \ref{theorem:second_order_closed_loop}}
\begin{proof}
Assume that $v_i^\star(\cdot|t_k)$ and $z_i^\star(\cdot|t_k)$ are the optimal solutions at $t_k$.
A feasible solution at $t_{k+1}$ can be constructed with a two-step control strategy 
\begin{align}
	\scalebox{0.85}{$\displaystyle
		\bar v_i(\tau|t_{k+1}) =
		\begin{cases}
			v_i^\star(\tau|t_k),
			\qquad \qquad \qquad \qquad \tau \in [t_{k+1}, t_k+\mathcal{T}-1], \\[4pt]
			\frac{1}{\hat\theta_i}
			\Big[
			\frac{1}{\delta^2}
			\big(z_i^\star(t_k+\mathcal{T}|t_k)
			-\hat p_i^\star(t_k+\mathcal{T}|t_k)\big) \\
			\qquad
			-\frac{3}{2\delta}
			\hat{\dot p}_i^\star(t_k+\mathcal{T}|t_k)
			\Big],
			\qquad\qquad\qquad \tau=t_k+\mathcal{T}, \\[8pt]
			
			\frac{1}{\hat\theta_i}
			\Big[
			\frac{1}{\delta^2}
			\big(\hat p_i^\star(t_k+\mathcal{T}|t_k)
			-z_i^\star(t_k+\mathcal{T}|t_k)\big) \\
			\qquad
			+\frac{1}{2\delta}
			\hat{\dot p}_i^\star(t_k+\mathcal{T}|t_k)
			\Big],
			
			\qquad\qquad	\tau=t_k+\mathcal{T}+1, \\[8pt]
			0,
			\qquad \qquad \qquad \quad \tau \in [t_k+\mathcal{T}+2,
			t_{k+1}+\mathcal{T}-1],
		\end{cases}
		$}
	\label{eq:candidate_v_2nd}
\end{align}
\begin{align} 
	&\scalebox{0.9}{$\displaystyle
		\bar z_i(\tau|t_{k+1}) = 
		\begin{cases} 
			z_i^\star(\tau|t_k), 
			& \tau \in [t_{k+1}, t_k+\mathcal{T}], \\[4pt] 
			z_i^\star(t_k+\mathcal{T}|t_k), 
			& \tau \in [t_k+\mathcal{T}+1, 
			t_{k+1}+\mathcal{T}], 
		\end{cases}
		$}
	\label{eq:candidate_z_2nd} 
\end{align}
which gives
\begin{align}
	\scalebox{0.89}{$\displaystyle
		\bar{\hat p}_i(\tau|t_{k+1}) =
		\begin{cases}
			\begin{aligned}
				&\tfrac{1}{2}
				\hat p_i^\star(t_k+\mathcal{T}|t_k)
				+\tfrac{1}{2}
				z_i^\star(t_k+\mathcal{T}|t_k) \\
				&\quad
				+\tfrac{1}{4}\delta
				\hat{\dot p}_i^\star(t_k+\mathcal{T}|t_k),
			\end{aligned}
			\enspace\tau=t_k+\mathcal{T}+1, \\[4pt]
			z_i^\star(t_k+\mathcal{T}|t_k),
			\qquad \quad\tau\in[t_k+\mathcal{T}+2,
			t_{k+1}+\mathcal{T}].
		\end{cases}
		$}
	\label{eq:candidate_p_2nd}
\end{align}
\begin{align}
	\scalebox{0.91}{$\displaystyle
		\bar{\hat{\dot p}}_i(\tau|t_{k+1}) =
		\begin{cases}
			\begin{aligned}
				&-\tfrac{
					\hat p_i^\star(t_k+\mathcal{T}|t_k)
					-z_i^\star(t_k+\mathcal{T}|t_k)
				}{\delta} \\
				&\quad
				-\tfrac{1}{2}
				\hat{\dot p}_i^\star(t_k+\mathcal{T}|t_k),
			\end{aligned}
			\quad \quad \tau=t_k+\mathcal{T}+1, \\[4pt]
			0,
			\qquad\qquad\qquad\quad\enspace \tau\in
			[t_k+\mathcal{T}+2,
			t_{k+1}+\mathcal{T}].
		\end{cases}
		$}
	\label{eq:candidate_vel_2nd}
\end{align}
Under conditions \eqref{eq:second_order_state_condition}--\eqref{eq:second_order_velocity_condition}, the constructed solution satisfies all constraints in Problem~\ref{prob:general_robust_mpc}, hence it is feasible.
For the actual system, since $\tilde e_i(t_0)\in\mathcal E_i$, Lemma~\ref{lem:invariant_ellipsoid} guarantees $\tilde e_i(k)\in\mathcal E_i$ for all $k\geq t_0$. Then, following the same constraint-tightening argument as in the proof of Theorem~\ref{theorem:first_order_closed_loop}, the tightened constraints on $\hat p_i(k)$, $\hat{\dot p}_i(k)$, and $v_i(k)$ guarantee
$p_i(k)\in\mathbb P_i$, $\dot p_i(k)\in\mathbb V_i$, and $u_i(k)\in\mathbb U_i$.
Consider the Lyapunov function defined by
$\tilde W(t_k)
=
J_{t_k}(V^\star(\tau|t_k),Z^\star(\tau|t_k))
+
\tilde e(t_k)^T P\tilde e(t_k)$.
For the optimal nominal control input $v_i^\star(\tau|t_{k+1})$ and
consensus state $z_i^\star(\tau|t_{k+1})$ at time $t_{k+1}$, the analysis of
$\tilde W(t_{k+1})-\tilde W(t_k)$
is similar to that in the proof of Theorem~\ref{theorem:first_order_closed_loop}.
Substituting the feasible solution constructed in
\eqref{eq:candidate_v_2nd}--\eqref{eq:candidate_vel_2nd}
and applying Young's inequality to bound cross terms, we obtain
		{\small\begin{align*}
				&\tilde{W}(t_{k+1})-\tilde{W}(t_k)\\
				&\le \sum_{i=1}^N \Big\{
				(
				\tfrac{4r_i}{\hat\theta_i^2\delta^4}
				+\tfrac{3q_{xi}}{2}
				+\tfrac{2q_{vi}}{\delta^2}
				-p_{xi}
				)
				\|\hat p_i^\star(t_k+\mathcal{T}|t_k)\\
				&\qquad\qquad-z_i^\star(t_k+\mathcal{T}|t_k)\|^2+
				(
				\tfrac{5r_i}{\hat\theta_i^2\delta^2}
				+\tfrac{3q_{vi}}{2}
				+\tfrac{q_{xi}\delta^2}{8}
				-p_{vi}
				)\\
				&\qquad\qquad			\|\hat{\dot p}_i^\star(t_k+\mathcal{T}|t_k)\|^2-
				\sum_{\tau=t_k}^{t_{k+1}-1}
				\Big[
				q_{xi}\|\hat p_i^\star(\tau|t_k)-z_i^\star(\tau|t_k)\|^2\\
				&\qquad\qquad
				+q_{vi}\|\hat{\dot p}_i^\star(\tau|t_k)\|^2
				+r_i\|v_i^\star(\tau|t_k)\|^2
				\big]
				\Big\}\\
				&\qquad\qquad-c_e\sum_{\tau=t_k}^{t_{k+1}-1}\|\tilde e(\tau)\|^2
				+d_\omega\sum_{\tau=t_k}^{t_{k+1}-1}\|\omega(\tau)\|^2.
		\end{align*}}%
		Under \eqref{eq:second_order_weight_condition_1} and
		\eqref{eq:second_order_weight_condition_2}, the first two terms are
		non-positive. Let $\alpha_p=\min_i\{q_{xi}\}>0$ and
		$\alpha_v=\min_i\{q_{vi}\}>0$.
		Then, for any $\tau\in[t_k,t_{k+1})$, we obtain
		\begin{align} 
			&\scalebox{0.9}{$\displaystyle
				\tilde W(t_{k+1})-\tilde W(t_k) 
				\leq
				-\alpha_p 
				\|\hat p^\star(\tau|t_k)-z^\star(\tau|t_k)\|^2
				$}
			\label{eq:second_order_lyapunov_decrease}\\ 
			&\scalebox{0.9}{$\displaystyle
				\qquad\qquad\qquad\qquad\quad-\alpha_v 
				\|\hat{\dot p}^\star(\tau|t_k)\|^2 
				-c_e\|\tilde e(\tau)\|^2 
				+d_\omega\Delta t\,\bar\omega^2.
				$}
			\nonumber
		\end{align}
		Taking $\tau=t_k$ in
		\eqref{eq:second_order_lyapunov_decrease}, and using
		$
		\hat p^\star(t_k|t_k)=\hat p(t_k)$,
		$
		\hat{\dot p}^\star(t_k|t_k)=\hat{\dot p}(t_k)$,
		gives an ISS-Lyapunov inequality for the augmented state
		$
		\chi(t_k)
		\triangleq
		\operatorname{col}\{
		\hat p(t_k)-z^\star(t_k|t_k),
		\hat{\dot p}(t_k),
		\tilde e(t_k)
		\}.
		$
		Hence, by standard ISS arguments
		\cite{marruedo2002Inputtostate}, the augmented system is ISS at the
		MPC update instants. Since
		\(
		p(t_k)-z^\star(t_k|t_k)
		=
		\hat p(t_k)-z^\star(t_k|t_k)+e^p(t_k)
		\)
		and
		\(
		\|e^p(t_k)\|\leq\|\tilde e(t_k)\|,
		\)
		there exist $\bar\beta\in\mathcal{KL}$ and
		$\bar r\in\mathcal K$ such that
		$
		\|p(t_k)-z^\star(t_k|t_k)\| \leq
		\bar\beta(
		\|(\hat p(t_0)-z^\star(t_0|t_0),
		\hat{\dot p}(t_0),
		\tilde e(t_0))\|,
		t_k-t_0
		)
		+
		\bar r(\bar\omega)$.
		We next consider the behavior between two consecutive MPC update
		instants. Since $\tilde W(t_{k+1})\geq0$,
		\eqref{eq:second_order_lyapunov_decrease} implies that, for every
		$\tau\in[t_k,t_{k+1})$, there has
		\begin{align}
			&\alpha_p
			\|\hat p^\star(\tau|t_k)-z^\star(\tau|t_k)\|^2
			+\alpha_v
			\|\hat{\dot p}^\star(\tau|t_k)\|^2
			+c_e\|\tilde e(\tau)\|^2
			\nonumber\\
			&\qquad\leq
			\tilde W(t_k)
			+d_\omega\Delta t\,\bar\omega^2.
			\label{eq:second_order_interupdate_bound}
		\end{align}
		The ISS property established at the MPC update instants implies that
		$\tilde W(t_k)$ is ultimately bounded uniformly with respect to $k$.
		Since $\Delta t$ is uniformly bounded, the right-hand side of
		\eqref{eq:second_order_interupdate_bound} is therefore uniformly
		ultimately bounded for all $k\geq 0$.
		Consequently,
		$\hat p^\star(\tau|t_k)-z^\star(\tau|t_k)$ and
		$\tilde e(\tau)$ are uniformly ultimately bounded over all
		inter-update intervals. Moreover,
		considering
		$p(\tau)-z^\star(\tau|t_k)
		=
		\hat p^\star(\tau|t_k)-z^\star(\tau|t_k)
		+e^p(\tau)$ with
		$\|e^p(\tau)\|\leq\|\tilde e(\tau)\|$,
		it follows that
		\(
		\|p(\tau)-z^\star(\tau|t_k)\|
		\)
		remains uniformly ultimately bounded for every
		$\tau\in[t_k,t_{k+1})$ and every $k\geq0$.
		Finally, for any $\tau\in[t_k,t_{k+1})$, the consensus constraint
		\eqref{LZ} implies
		\(
		z_i^\star(\tau|t_k)
		=
		z_j^\star(\tau|t_k),
		\forall i,j.
		\)
		Therefore, we have
		$
		\|p_i(\tau)-p_j(\tau)\|
		=
		\|
		p_i(\tau)-z_i^\star(\tau|t_k)
		-
		\bigl(p_j(\tau)-z_j^\star(\tau|t_k)\bigr)
		\|
		\leq
		\|p_i(\tau)-z_i^\star(\tau|t_k)\|
		+
		\|p_j(\tau)-z_j^\star(\tau|t_k)\|,
		$
		which completes the proof.
\end{proof}

\bibliographystyle{IEEEtran}
\bibliography{IEEEabrv,Bibliography/bibtex}

\end{document}